\documentclass[
 aip,
 amsmath,amssymb,
preprint
]{revtex4-2}

\usepackage{natbib}
\usepackage{graphicx}
\usepackage{dcolumn}
\usepackage{bm}
\usepackage{multirow}
\usepackage{amsmath}
\usepackage{makecell}
\usepackage{tikz}

\usepackage[utf8]{inputenc}
\usepackage[T1]{fontenc}
\usepackage{mathptmx}

\begin{document}

\preprint{}

\title{High-fidelity reconstruction of turbulent flow from spatially limited data using enhanced super-resolution generative adversarial network}

\author{Mustafa Z. Yousif}
\author{Linqi Yu}

\author{Hee-Chang Lim}
\email[]{Corresponding author, hclim@pusan.ac.kr}
\thanks{}
\affiliation{School of Mechanical Engineering, Pusan National University, 2, Busandaehak-ro 63beon-gil, Geumjeong-gu, Busan, 46241, Rep. of KOREA}


\begin{abstract}
In this study, a deep learning-based approach is applied with the aim of reconstructing high-resolution turbulent flow fields using minimal flow fields data. A multi-scale enhanced super-resolution generative adversarial network with a physics-based loss function is introduced as a model to reconstruct the high-resolution flow fields. The model capability to reconstruct high-resolution laminar flows is examined using data of laminar flow around a square cylinder. The results reveal that the model can accurately reproduce the high-resolution flow fields even when limited spatial information is provided. The case of turbulent channel flow is used to assess the ability of the model to reconstruct the high-resolution wall-bounded turbulent flow fields. The instantaneous and statistical results obtained from the model agree well with the ground truth data, indicating that the model can successfully learn to map the coarse flow fields to the high-resolution ones. Furthermore, the computational cost of the proposed model, which is examined carefully, is found to be effectively low. This demonstrates that using high-fidelity training data with physics-guided generative adversarial network-based models can be practically efficient in reconstructing high-resolution turbulent flow fields from extremely coarse data.

\end{abstract}

\maketitle

\section{Introduction}\label{sec:introduction}
The reconstruction of flow fields using spatially limited data has long been a topic of interest in the fluid dynamics community. With the rapid development of particle image velocimetry (PIV) and computational power for performing direct numerical simulations (DNS), high-fidelity turbulence data can be generated. However, as turbulence exhibits a chaotic behaviour with a wide range of spatio-temporal scales, an expensive experimental setup and a substantial computational cost are required to obtain high-resolution turbulent flow fields. With the availability of enormous amounts of data that can be obtained from experimental and DNS studies, deep learning techniques have a great potential to serve as an alternative data-driven methods to tackle turbulent flow problems. Deep learning is a subset of machine learning, where deep neural networks are used for classification, prediction, and feature extraction \cite{LeCunetal2015}. In this study, we demonstrate a deep learning-based approach to reconstruct high-resolution laminar and turbulent flow fields using coarse data, which are represented by a limited number of distributed points. \par

Recently, there have been considerable developments in deep learning algorithms that can be practically utilised in the field of fluid dynamics \cite{Bruntonetal2020, Kutz2017}. Various deep learning-based approaches have been applied for different applications in turbulence, such as turbulence modelling \cite{Duraisamyetal2019, Gamahara&Hattori2017, Lingetal2016, Wangetal2017}, flow prediction \cite{Lee&You2019, Srinivasanetal2019} and flow control \cite{Fanetal2020, Rabaultetal2019}. On the other hand, high-resolution turbulent flow reconstruction has recently become an active research topic after the introduction of various supervised and unsupervised deep learning-based algorithms designed to deal with high-resolution image reconstruction \cite{Bashiretal2021}. These developments in deep learning techniques accompanied with the rapid growth in graphics processing unit (GPU) power have opened the doors to explore novel methods that could help reconstruct super-resolution flow fields using extremely low-resolution experimental measurements or simulation results rather than applying traditional handcrafted super-resolution methods, such as bicubic interpolation \cite{Keys1981}. \par

Fukami {\it et al.} \cite{Fukami2019, Fukami2021} proposed models based on convolutional neural networks (CNNs) for spatial and spatio-temporal super-resolution reconstruction of turbulent flows. They reported good reconstruction of velocity and vorticity fields using extremely low-resolution data along with good prediction of the temporal evolution for the intervals for which the model was trained. 

Onishi {\it et al.} \cite{Onishietal2019} presented a super-resolution method based on CNN for reconstructing high-resolution data from low-resolution urban meteorology simulation data. Their results revealed that the CNN-based model highly outperformed the conventional interpulation methods. Liu et al. \cite{Liuetal2020} considered the temporal behaviour of the flow in the reconstruction of the high-resolution turbulent flow fields using CNN-based model. They showed that by applying this approach, the reconstruction accuracy could be remarkably improved compared with the static CNN-based model. \par

Recently, Kim et al. \cite{Kimetal2021} showed that unsupervised deep learning has great potential for reconstructing high-resolution turbulence using unpaired training data. They used a cycle-consistent generative adversarial network (CycleGAN) \cite{Zhuetal2017} to reconstruct  high-resolution velocity fields from low-resolution DNS and large eddy simulation (LES) data. Their results showed better reconstruction accuracy compared with that of bicubic interpolation and CNN-based model.  \par
 
In terms of experimental studies, Deng {\it et al.} \cite{Dengetal2019} applied a super-resolution GAN (SRGAN) \cite{Ledigetal2017} and enhanced SRGAN (ESRGAN) \cite{Wangetal2018} to reconstruct high-resolution flow fields using PIV measurements of flow around a cylinder. They reported an accurate reconstruction of the mean and fluctuation flow fields. They observed that the reconstruction capability of ESRGAN was better than that of SRGAN. Cai {\it et al.} \cite{Caietal2019} proposed a CNN-based model for estimating high-resolution velocity fields from PIV measurments. Their model showed a better performance compared with the traditional cross-correlation algorithms. Morimoto {\it et al.} \cite{Morimotoetal2020} applied a CNN-based model to estimate the velocity fields using PIV measurements with missing regions. They reported a relatively good estimation of the missing regions in the velocity.\par

In this study, we focus on the reconstruction of high-resolution flow fields using extremely spatially limited data that are represented by distributed points in the flow. This approach can mimic high-resolution flow fields reconstruction from spatially limited PIV or hot wire measurements. We apply ESRGAN-based model, i.e. multi-scale ESRGAN (MS-ESRGAN), to reconstruct high-resolution flow fields using data with various coarseness levels. As representative examples, we consider the DNS of laminar flow around a square cylinder and turbulent channel flow.  \par

The remainder of this paper is organised as follows. In Section 2, the methodology of reconstruction high-resolution flow fields using the proposed deep learning model is explained. The generation of the training data using DNS is described in Section 3. In Section 4, the results of testing the proposed model are discussed. Finally, the conclusions of this study are presented in Section 5. \par

\section{Methodology}

Since the first version of GAN was introduced by Goodfellow {\it et al.} \cite{Goodfellowetal2014}, variants of GAN have been proposed to tackle different types of image transformation and super-resolution problems \cite{Ledigetal2017, Mirza&Osindero2014, Wangetal2018, Zhuetal2017}. The architecture of GAN is designed to be different from the traditional architecture of multilayer perceptron (MLP) or CNN-based models. In GAN, two adversarial networks, i.e. the generator ($G$) and the discriminator ($D$), compete with each other. Here, $G$ generates fake images similar to the real ones, whereas $D$ distinguishes the fake images from the real ones. $G$ and $D$ are usually MLPs or CNNs that are trained simultaneously. The goal of the training process is to make $G$ generate fake images that are difficult to distinguish using $D$. This process can be expressed as a min-max two-player game with a value function $V(D, G)$ such that:

\begin{equation} \label{eqn:eq1}
\begin{split}
\substack{min\\G} ~\substack{max\\D} ~V(D,G) = \mathbb{E}_{x_r \sim P_{data(x_r)}} [ {\rm log} D(x_r )] + \mathbb{E}_{z \sim P_z(z) } [ {\rm log} (1-D(G(z)))],
\end{split}
\end{equation}

\noindent where $x_r$ is the image from the ground truth data, whereas $P_{data(x_r )}$ is the real image distribution. $\mathbb{E}$ represents the operation of calculating the average of all the data in the training mini-batch. In the second right term of Eq.~\ref{eqn:eq1}, $z$ is a random vector used as an input to $G$, whereas $D(x_r )$ represents the probability that the image is real and not generated by $G$. $G(z)$ is the output from $G$, which is expected to generate an image that is similar to the real image, such that the value of $D(G(z))$ is close to 1. On the other hand, in $D$, $D(x_r )$ returns a value close to 1, whereas $D(G(z))$ returns a value close to 0. Thus, in the training process, $G$ is trained in a direction that minimises $V(D,G)$, and $D$ is trained in a direction that maximises $V(D,G)$. After successful training, $G$ is expected to produce an image with a distribution similar to the real image that $D$ cannot judge whether it is real or fake.

This study applies a newly developed high-fidelity deep learning framework based on ESRGAN \cite{Wangetal2018} to reconstruct a high-resolution flow fields using extremely coarse data as input to $G$. We adopted the generator network to obtain MS-ESRGAN. The architecture of $G$ in MS-ESRGAN is shown in Fig.~\ref{eqn:eq1}(a). Here, $G$ consists of a deep convolution neural network represented by residuals in residual dense blocks (RRDBs) and multi-scale branches. 

The coarse input data are first passed through a convolution layer and then through a series of RRDBs. The multi-scale part, which consists of three parallel convolutional sub-models with different kernel sizes, is applied to the data features that are extracted by the RRDBs. Finally, the outputs of the three branches are simply summed and passed through a final convolutional layer to generate a high-resolution fake image ($x_f$). Fig.~\ref{eqn:eq1}(b) shows the architecture of $D$. As mentioned earlier, $D$ is designed to distinguish between the fake high-resolution image and the real image. The fake and real images are fed to $D$ and passed through a series of convolutional, batch normalisation, and leaky ReLU layers. Then, the data are passed through a final convolutional layer. The non-transformed discriminator outputs using the real and fake images, i.e. $C(x_r )$ and $C(x_f )$, are used to calculate the relativistic average discriminator value $D_{Ra}$ \cite{Jolicoeur-Martineau2018}:

\begin{equation} \label{eqn:eq2}
D_{Ra} (x_r , x_f ) = \sigma (C \left(x_r ) \right) - \mathbb{E}_{x_f} \left[C ( x_f) \right],
\end{equation}

\begin{equation} \label{eqn:eq3}
D_{Ra} (x_f , x_r ) = \sigma (C \left(x_f ) \right) - \mathbb{E}_{x_r} \left[C ( x_r) \right],
\end{equation}

\noindent where $\sigma$ is the sigmoid function. In Eqs.~\ref{eqn:eq2} and ~\ref{eqn:eq3}, $D_{Ra}$ predicts the probability that the output from $D$ using the real image is relatively more realistic than the output using the fake image.
The discriminator loss is then defined as:

\begin{equation} \label{eqn:eq4}
L_D^{Ra} = -\mathbb{E}_{x_r} \left[ {\rm log} (D_{Ra} (x_r , x_f )) \right] - \mathbb{E}_{x_f} \left[ {\rm log} (1 - D_{Ra} (x_f , x_r )) \right].
\end{equation}

The adversarial loss of the generator can be expressed in a symmetrical form as:

\begin{equation} \label{eqn:eq5}
L_G^{Ra} = -\mathbb{E}_{x_r} \left[ {\rm log} (1 - D_{Ra} (x_r , x_f )) \right] - \mathbb{E}_{x_f} \left[ {\rm log} (D_{Ra} (x_f , x_r )) \right].
\end{equation}

In addition to the adversarial loss, four additional loss terms are used to form the combined loss function of the generator: pixel loss $(L_{pixel})$, perceptual loss $(L_{perceptual})$, gradient loss $(L_{gradient})$, and Reynolds stress loss $( L_{Reynolds~stress})$. $L_{pixel}$ is the pixel-based error between the generated data and the ground truth data. $L_{perceptual}$ represents the difference in the extracted features of the real and fake data. The pre-trained CNN VGG-19 \cite{Simonyan&Zisserman2015} is used to extract the features. While one layer of VGG-19 was applied in the model of Wang {\it et al.} \cite{Wangetal2018}, we apply three different layers to extract the features. This strategy has showed a remarkable improvement in the training stability. $L_{gradient}$ represents the difference in the gradient of the generated fake data and real data, whereas $L_{Reynolds~stress}$ is the error that represents the difference between the Reynolds stress tensor of the velocity data that are obtained from the generator and the Reynolds stress tensor of the ground truth velocity data. The mean squared error (MSE) is used to calculate all the loss terms except $L_G^{Ra}$. \par 

The combined loss function of the generator is expressed as: 

\begin{equation} \label{eqn:eq6}
\mathcal{L}_G = L_G^{Ra} + \lambda_1 L_{pixel} + L_{perceptual} + \lambda_2 L_{gradient} + \lambda_3 L_{Reynolds~stress},     
\end{equation}

\noindent where $\lambda_1, \lambda_2$, and $\lambda_3$ represent the coefficients that are used to balance the different loss terms whose values are set to be 5000, 10, and 100, respectively. $L_{gradient}$, is used to consider the non-uniform distribution of the grid points in the training process, whereas $L_{perceptual}$ can help in overcoming  the training instability. The turbulence statistics can be improved by applying $L_{Reynolds~stress}$, which forces the model to consider the components of the Reynolds stress tensor in the training process. \par

In this study, the adaptive moment estimation (ADAM) optimisation algorithm \cite{Kingma&Ba2017} is applied to update the weights of the model. The training data are divided into mini-batches, and the size of each mini-batch is set to be 16. \par

\begin{figure}
\centering 
\includegraphics[angle=0, trim=0 0 0 0, width=0.9\textwidth]{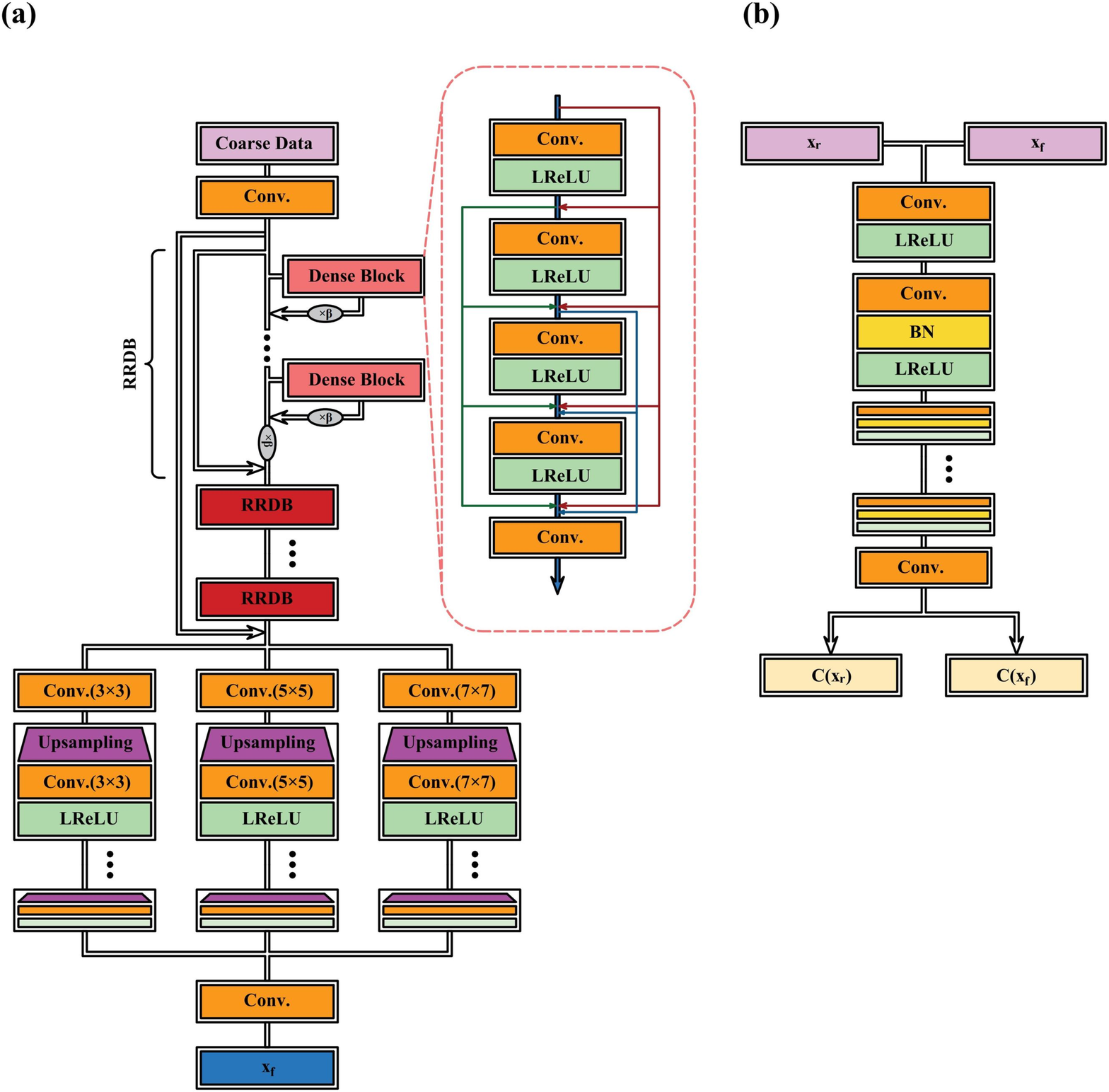}
\caption[]{MS-ESRGAN architecture: (a) the generator ($\beta$ is the residual scaling parameter = 0.2), and (b) the discriminator.}
\label{fig:1-ESRGAN}
\end{figure}

\section{Generation of training data}

Two examples are used in this study, two-dimensional laminar flow around a square cylinder at $Re_d = 100$ is used as a demonstration, and turbulent channel flow at $Re_\tau = 180$ is used as a test case for reconstructing high-resolution wall-bounded turbulent flow. The training data of each example are generated by performing DNS. \par

The momentum equation for an incompressible viscous fluid is:

\begin{equation}\label{eqn:eq7}
\frac{\partial {\bf u}}{\partial t} + {\bf u}\cdot \nabla {\bf u} = - \frac{1}{\rho} \nabla p + {\it \nu} \nabla^2 {\bf u},
\end{equation}

\noindent where ${\bf u}$ is the velocity of the fluid, $\rho$ is the density, $p$ is the pressure, and $\nu$ is the kinematic viscosity. The continuity equation that expresses the incompressibility of the fluid is defined as:

\begin{equation}\label{eqn:eq8}
\nabla \cdot {\bf u} = 0.
\end{equation}

The open-source computational fluid
dynamics (CFD)  finite-volume code OpenFOAM-5.0x is used to perform the DNS. In the case of two-dimensional laminar flow around a square cylinder, the Reynolds number is based on the free-stream velocity and cylinder width, i.e. $Re_d = U_\infty d /{\it \nu}$, where $U_\infty$ is the free-stream velocity and $d$ is the cylinder width. The domain size is set to be $(xd{\times}yd) = (15{\times}20)$ with the corresponding grid size of $(381{\times}221)$. The time step of the simulation is set to be $\Delta t =10^{-2}$. The statistics obtained from the simulation have been validated agnainst DNS results obtained by Anzai {\it et al.}\cite{Anzaietal2017}  

In the case of turbulent channel flow, the friction Reynolds number, i.e. $Re_\tau = u_\tau \delta / {\it \nu}$ is set to be 180, where $u_\tau$ is the friction velocity and $\delta$ is half of the channel height. The dimensions of the computational domain are set to be $4\pi \delta$, $2\delta$, and $2 \pi \delta$ in the streamwise ($x$), wall-normal ($y$), and spanwise ($z$) directions, respectively. The corresponding grid points are 256, 128, and 256, respectively. Uniform grid points distributions with spaceing $\Delta x^+$ $\approx$ 6.3 and $\Delta z^+$ $\approx$ 2.8 are used in the streamwise and spanwise directions. Note that the superscript $+$ indicates that the quantity is made dimensionless using the wall variables, i.e. $u_\tau$ and $\it \nu$. A non-uniform grid points distribution is used in the wall-normal direction. The first grid piont away from the wall is located at $y^+$ $\approx$ 0.63 and the maximum spacing (at the centreline of the channel), i.e. $\Delta y^+_{max}$ $\approx$ 6.4. The periodic boundary condition is assigned to the streamwise and spanwise directions, whereas the no-slip condition is applied to the upper and lower walls of the channel. The time step of the simulation is set to be $\Delta t = 10^{-2}$ corresponding to $\Delta t^+ = 0.1134$. The turbulent statistics obtained from the simulation have been validated using DNS data obtained by Moser {\it et al.} \cite{Moseretal1999}.\par

For both simulations, the pressure implicit split operator algorithm is employed to solve the coupled pressure momentum system. The convective fluxes are discretised with a second-order accurate linear upwind scheme and all other discretisation schemes that are used in each simulation have second-order accuracy. The maximum Courant–Friedrichs–Lewy (CFL) number is maintained to be less than 1 to ensure simulation stability.\par

The training data are obtained with 6,000 snapshots from the simulation of two-dimensional laminar flow around a square cylinder and 10,000 snapshots from a single ($y-z$) plane of the turbulent channel simulation. The domain size used for training in the case of two-dimensional laminar flow around square cylinder is fixed to be $(xd \times yd) = (16.82 \times 8)$, which is equivalent to a grid size of ($320 \times 160$). The same grid size obtained from the simulation is used for training in the turbulent channel flow case. In both cases, the interval between the collected snapshots of the flow fields is 10 times the time step used in the simulation.\par

\begin{figure}
\centering 
\includegraphics[angle=0, trim=0 0 0 0, width=0.9\textwidth]{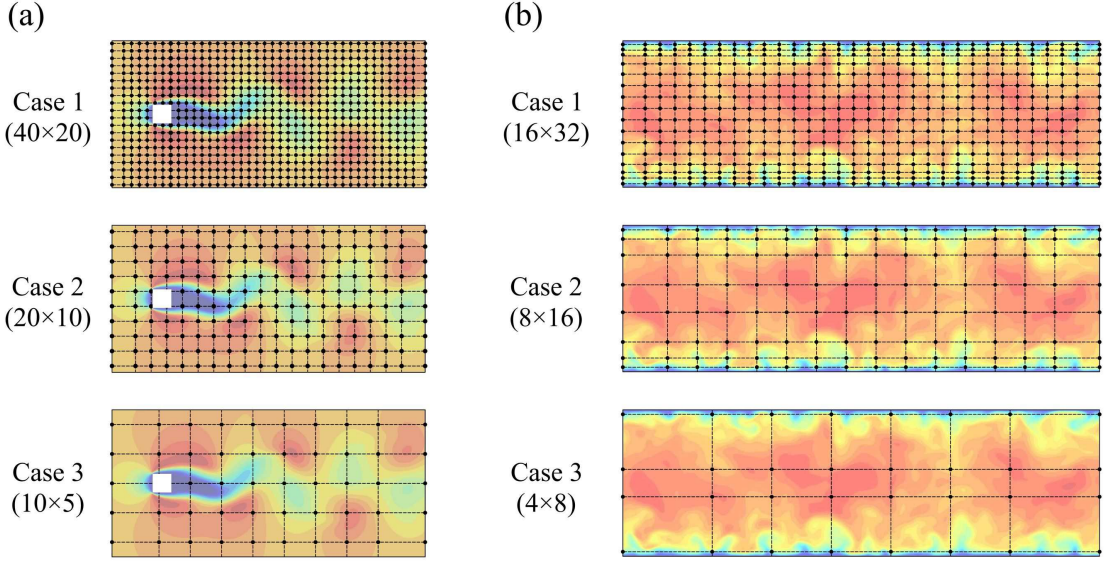}
\caption[]{The distribution of the selected points at three different coarseness levels for the case of (a) two-dimensional laminar flow around a square cylinder, and (b) turbulent channel flow.} 
\label{fig:2-Coarseness}
\end{figure}

As mentioned earlier, the low-resolution data are obtained by selecting a distributed points in the flow, i.e. no filtering operation is used. The distribution of the selected points at different coarseness levels for the two cases is shown in Fig.~\ref{fig:2-Coarseness}. The distribution of the points ($n_x{\times}n_y$) in the case  of two-dimensional laminar flow around a square cylinder, as shown in Fig.~\ref{fig:2-Coarseness}(a), has three levels of coarseness: case 1 (40 $\times$ 20), case 2 (20 $\times$ 10), and case 3 (10 $\times$ 5). Figure~\ref{fig:2-Coarseness}(b) shows the three coarseness levels of the points distribution ($n_y \times n_z$) in the case of turbulent channel flow: case 1 (16 $\times$ 32), case 2 (8 $\times$ 16), and case 3 (4 $\times$ 8). To prepare the data for the training process, the data are normalised using the min-max normalisation function to produce values between 0 and 1. The shape of the input data to $G$ is fixed to be (40 $\times$ 20) in the case of laminar flow around a square cylinder and (16$\times$32) in the case of turbulent channel flow. To achieve these shapes, upsampling is performed on cases 1 and 2 for each of the two cases used in this study. \par

\section{Results and discussion}

\subsection{Flow around a square cylinder}

\begin{figure}
\centering 
\includegraphics[angle=0, trim=0 0 0 0, width=0.7\textwidth]{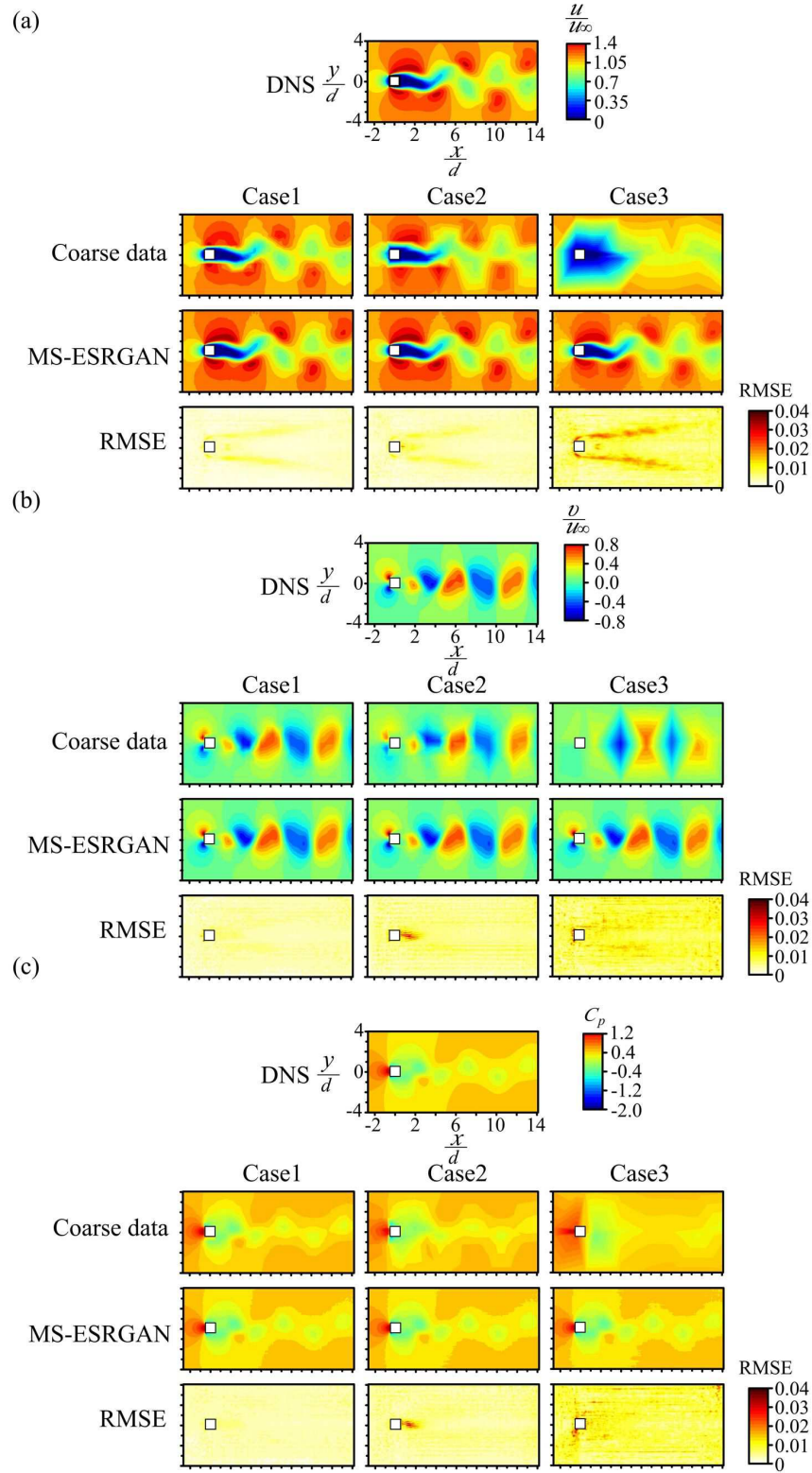}
\caption[]{Reconstructed instantaneous flow fields and RMSE for the case of two-dimensional laminar flow around a square cylinder: (a) streamwise velocity, (b) spanwise velocity, and (c) pressure.} 
\label{fig:3-Cyl-RMSE}
\end{figure}

In this section, we examine the ability of MS-ESRGAN to reconstruct high-resolution flow fields using coarse data of two-dimensional laminar flow around a square cylinder. Note that all the results are obtained using the test data that are not included in the training process. The reconstructed instantaneous velocity fields ($u$ and $\upsilon$), pressure field, and root-mean-square error (RMSE) of the reconstruction are shown in Fig.~\ref{fig:3-Cyl-RMSE}. Here, the velocity components are normalised by $U_\infty$, and the dimensionless pressure is given as $C_p = (p- p_\infty) / 0.5\rho U_\infty $, where  $p_\infty$ is the free-stream pressure. The reconstructed fields show a commendable agreement with the DNS results even when a high coarseness level is used (i.e. case 3). As shown in the figure, the RMSE is proportional to the coarseness level with a well-accepted maximum value.

\begin{figure}
\centering 
\includegraphics[angle=0, trim=0 0 0 0, width=0.8\textwidth]{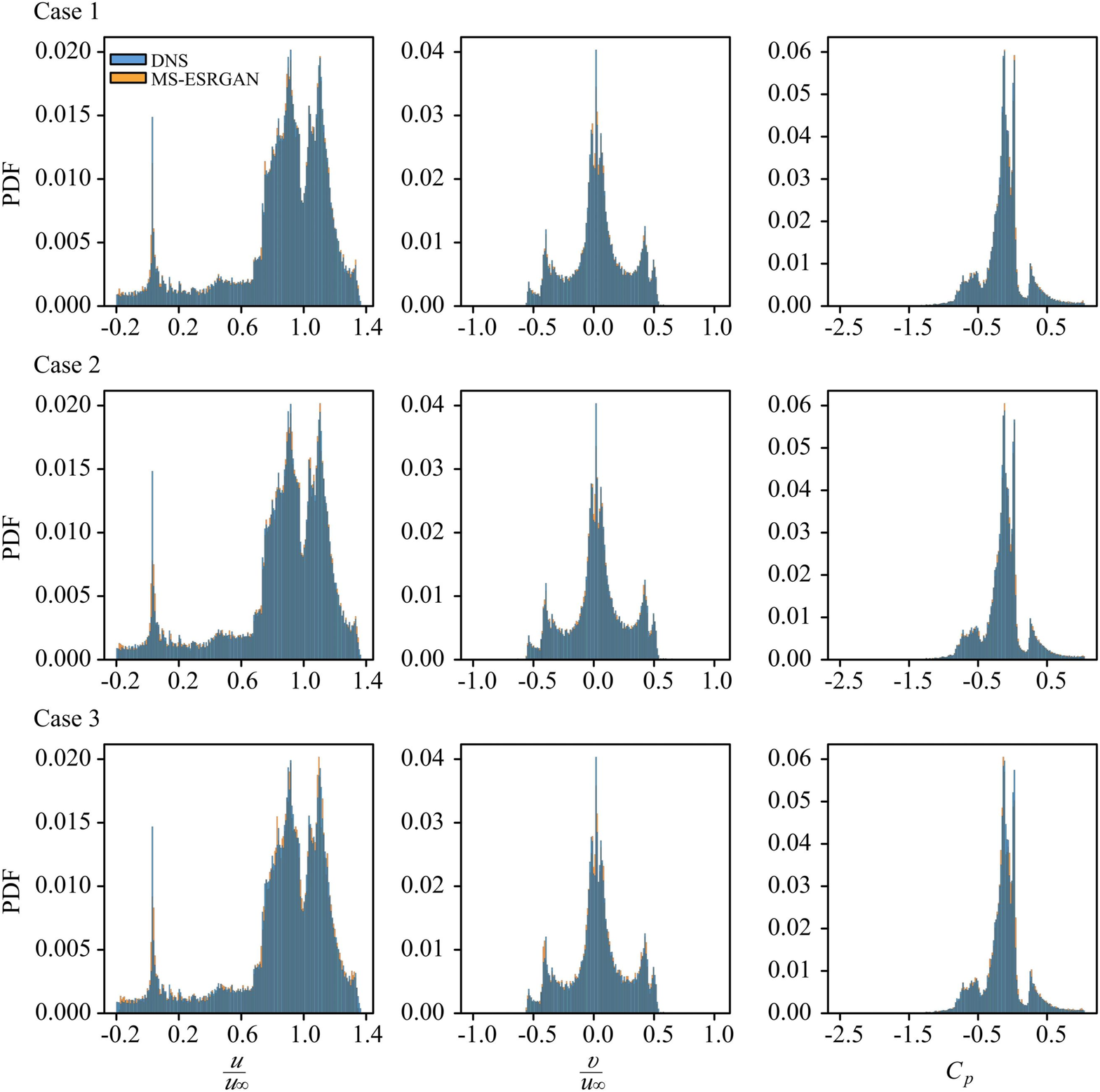}
\caption[]{Probability density functions of the reconstructed velocity components and pressure for the case of two-dimensional laminar flow around a square cylinder.} 
\label{fig:4-Cyl-PDF}
\end{figure}

To further validate the model, the capability of the model to reconstruct high-resolution flow fields is examined statistically. Figure~\ref{fig:4-Cyl-PDF} shows the probability density function (PDF) of the reconstructed velocity components and the pressure. All the reconstruction results show excellent agreement with the results obtained from the DNS, thus indicating the ability of the model to reconstruct instantaneous high-resolution flow fields. \par

\begin{figure}
\centering 
\includegraphics[angle=0, trim=0 0 0 0, width=0.9\textwidth]{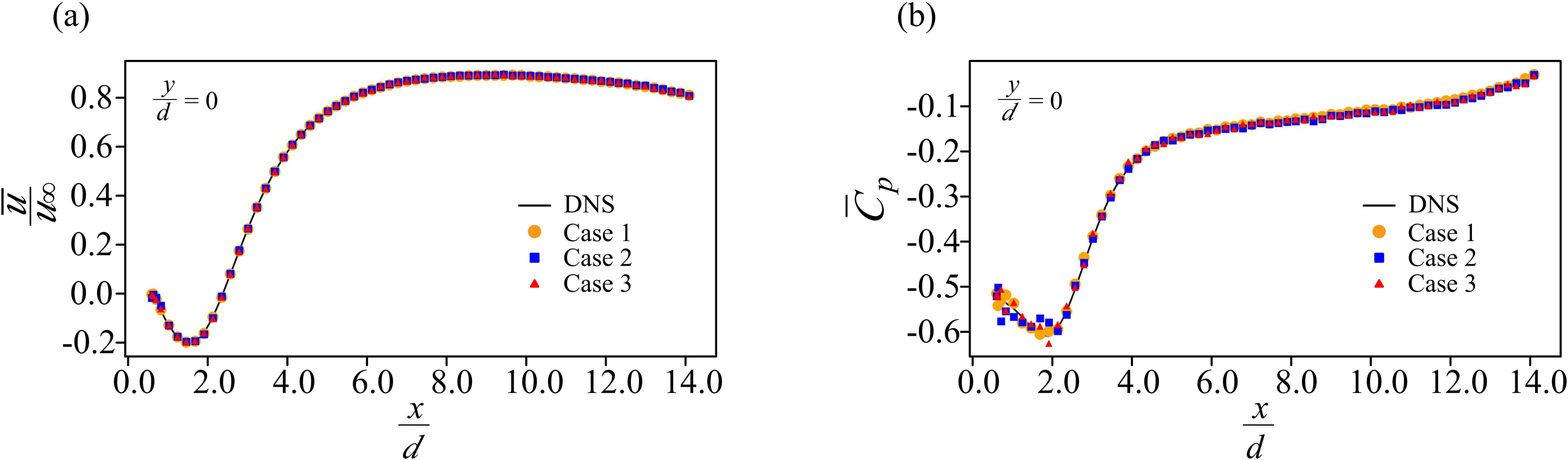}
\caption[]{Profiles of the reconstructed mean streamwise velocity (a) and mean pressure (b) for the case of two-dimensional laminar flow around a square cylinder.} 
\label{fig:5-Cyl-MeanVP}
\end{figure}

To observe the flow characteristics, the profiles of the mean streamwise velocity and mean pressure are calculated using 5000 reconstructed snapshots. As shown in Fig.~\ref{fig:5-Cyl-MeanVP}(a) and (b), the mean streamwise velocity and pressure are in commendable agreement with the results obtained from the DNS. However, a scattering in the pressure values near the rear of the cylinder, i.e. $x/d$ $\approx$ $0.5$ can be seen in Fig.~\ref{fig:5-Cyl-MeanVP}(b) while it is not shown in the velocity profile in Fig.~\ref{fig:5-Cyl-MeanVP}(a). This behaviour can be attributed to the rapid change of the pressure values in this region and the fewer physics constraints in the combined loss function that are dedicated to the pressure compared with those that are dedicated to the velocity components. For the pressure, there is only the gradient term, while for the velocity, there are the gradient and the Reynolds stress tenser terms. We believe that to achieve more accurate reconstruction for the pressure field, more studies that focus on the pressure-based physics constrains are required.

\begin{figure}
\centering 
\includegraphics[angle=0, trim=0 0 0 0, width=1.0\textwidth]{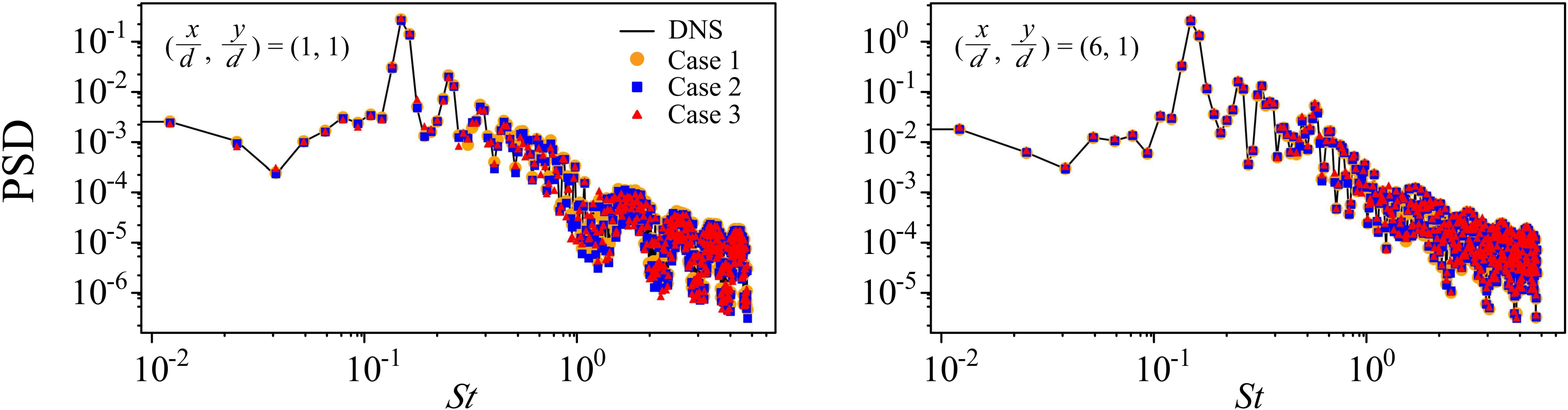}
\caption[]{Power spectrum density of the reconstructed streamwise velocity fluctuations at two different streamwise locations for the case of two-dimensional laminar flow around a square cylinder.} 
\label{fig:6-Cyl-Spectrum}
\end{figure}

Figure~\ref{fig:6-Cyl-Spectrum} shows the power spectrum density (PSD) of the streamwise velocity fluctuations at two different streamwise locations plotted against Strouhal number ($St$=$fd$/$U_\infty$), where $f$ is the frequency. The results obtained from the reconstructed velocity data are in excellent
agreement with the results obtained from the DNS for all the three levels of coarseness, indicating that the reconstructed data match the temporal behaviour of the ground truth data accurately. \par

The aforementioned results suggest that MS-ESRGAN can reconstruct laminar flow fields with high spatial resolution and reproduce the same dynamics as the ground truth data. \par

\subsection{Turbulent channel flow}

\begin{figure}
\centering 
\includegraphics[angle=0, trim=0 0 0 0, width=0.8\textwidth]{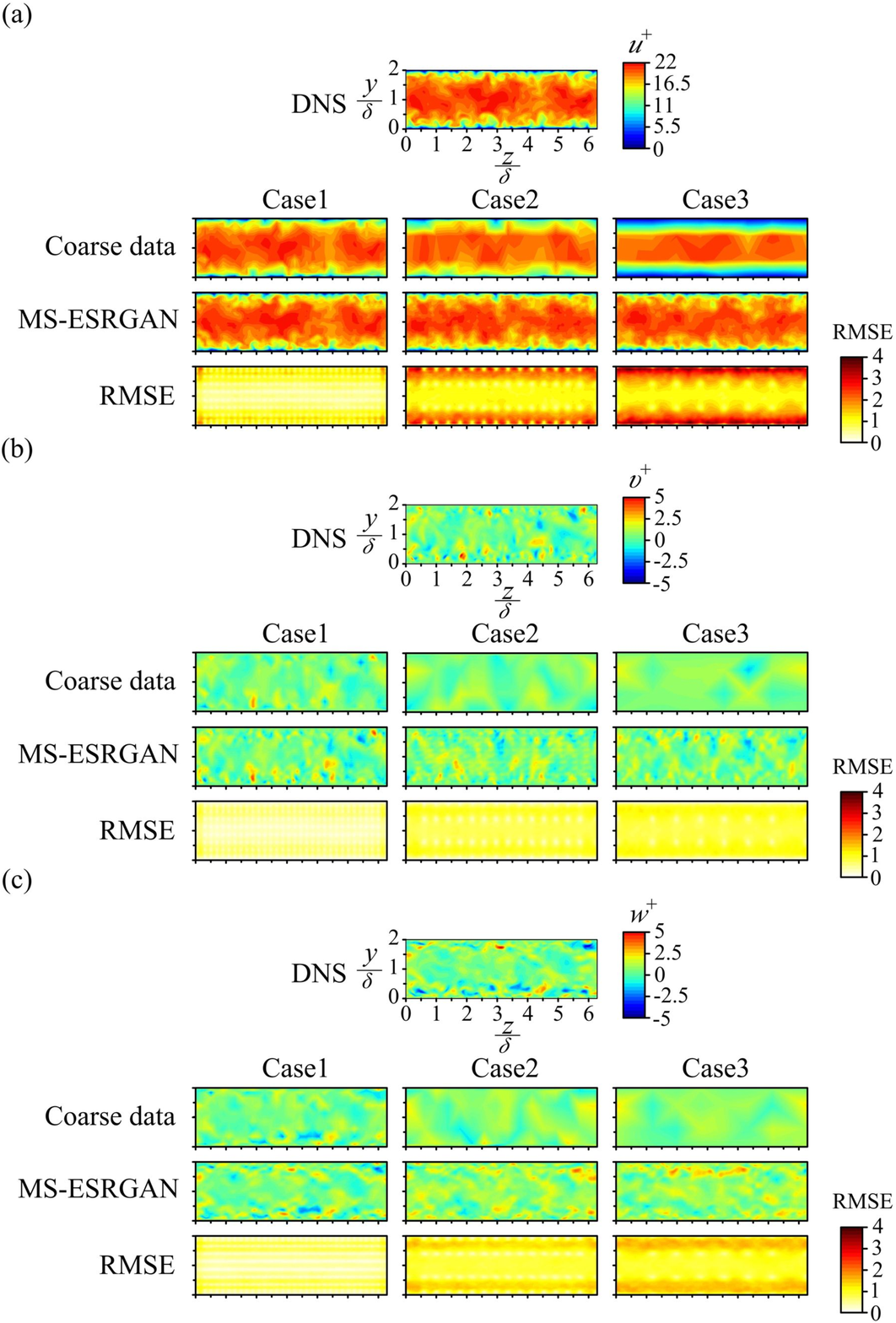}
\caption[]{Reconstructed instantaneous velocity fields and RMSE for the case of turbulent channel flow: (a) streamwise velocity, (b) wall-normal velocity, and (c) spanwise velocity.} 
\label{fig:7-Chan-RMSE}
\end{figure}

The ability of MS-ESRGAN to reconstruct high-resolution wall-bounded turbulent flow fields is validated in this section using a plane normal to the streamwise direction in the turbulent channel flow case, i.e. ($y-z$) plane. The reconstructed instantaneous velocity fields (${u^+}$, ${\upsilon^+}$, and ${w^+}$), and RMSE are shown in Fig.~\ref{fig:7-Chan-RMSE}. The results of all the velocity components are in agreement with the DNS data, including that of case 3, where minimal information about the flow field is available. While the RMSE of the streamwise velocity component is noticeably affected by the level of coarseness, the wall-normal and spanwise velocity components show less sensitivity to the coarseness level.\par

\begin{figure}
\centering 
\includegraphics[angle=0, trim=0 0 0 0, width=0.8\textwidth]{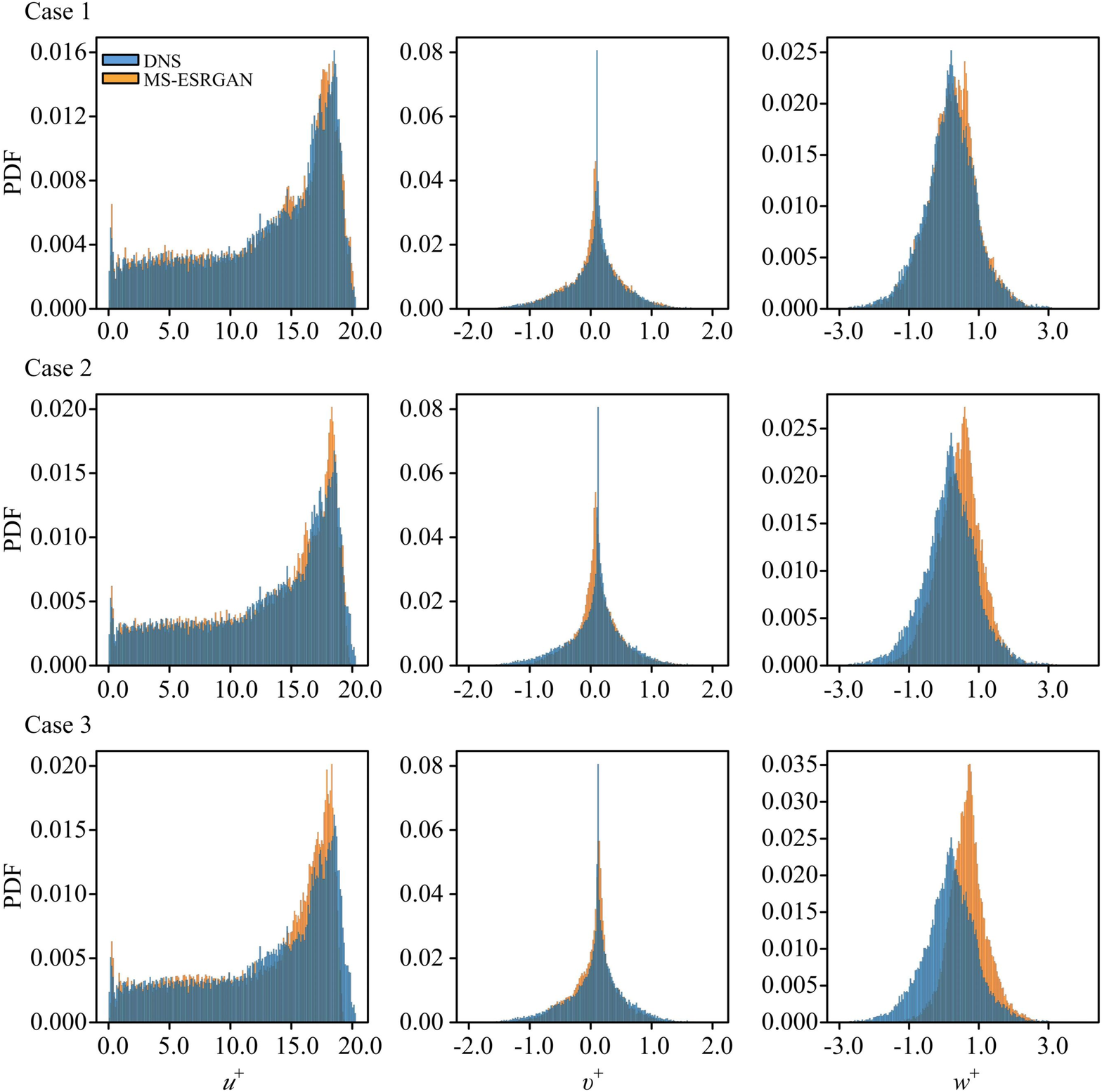}
\caption[]{Probability density functions of the reconstructed velocity components for the case of turbulent channel flow.} 
\label{fig:8-Chan-PDF}
\end{figure}

As shown in Fig.~\ref{fig:8-Chan-PDF}, the PDF plots of the reconstructed velocity components reveal a commendable agreement with the results of the streamwise and wall-normal velocities that obtained from the DNS. However, a slight deviation can be observed for the spanwise velocity, which increases with the increase in coarseness level. This can be attributed to the limited information about the spanwise velocity component that is available considering the more random behaviour of the spanwise velocity component compared with that of the other two velocity components.

\begin{figure}
\centering 
\includegraphics[angle=0, trim=0 0 0 0, width=0.8\textwidth]{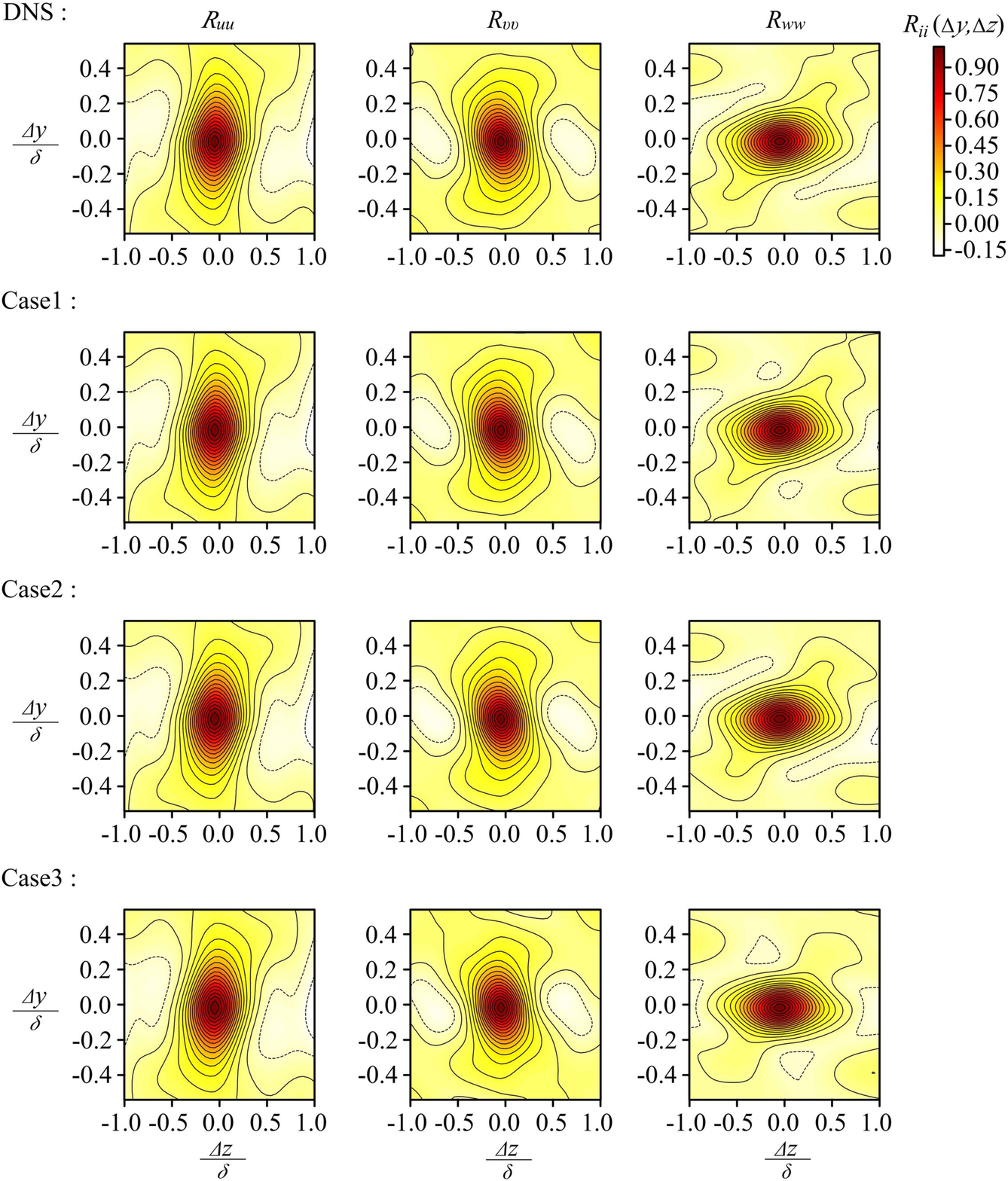}
\caption[]{Two-dimensional cross-correlations of the reconstructed velocity components for the case of turbulent channel flow.} 
\label{fig:9-2D-Corrln}
\end{figure}

To further examine the capability of the model to reproduce the velocity fields with accurate spatial resolution, two-dimensional cross-correlation $(R_{ii}(\Delta y,\Delta z ))$ plots of the velocity fluctuations are examined, as shown in Fig.~\ref{fig:9-2D-Corrln}. The correlations of all the three levels of coarseness are generally in good agreement with the correlations obtained from the DNS results, indicating the excellent ability of  MS-ESRGAN to reproduce the high-resolution velocity fields with an accurate spatial distribution.

\begin{figure}
\centering 
\includegraphics[angle=0, trim=0 0 0 0, width=0.9\textwidth]{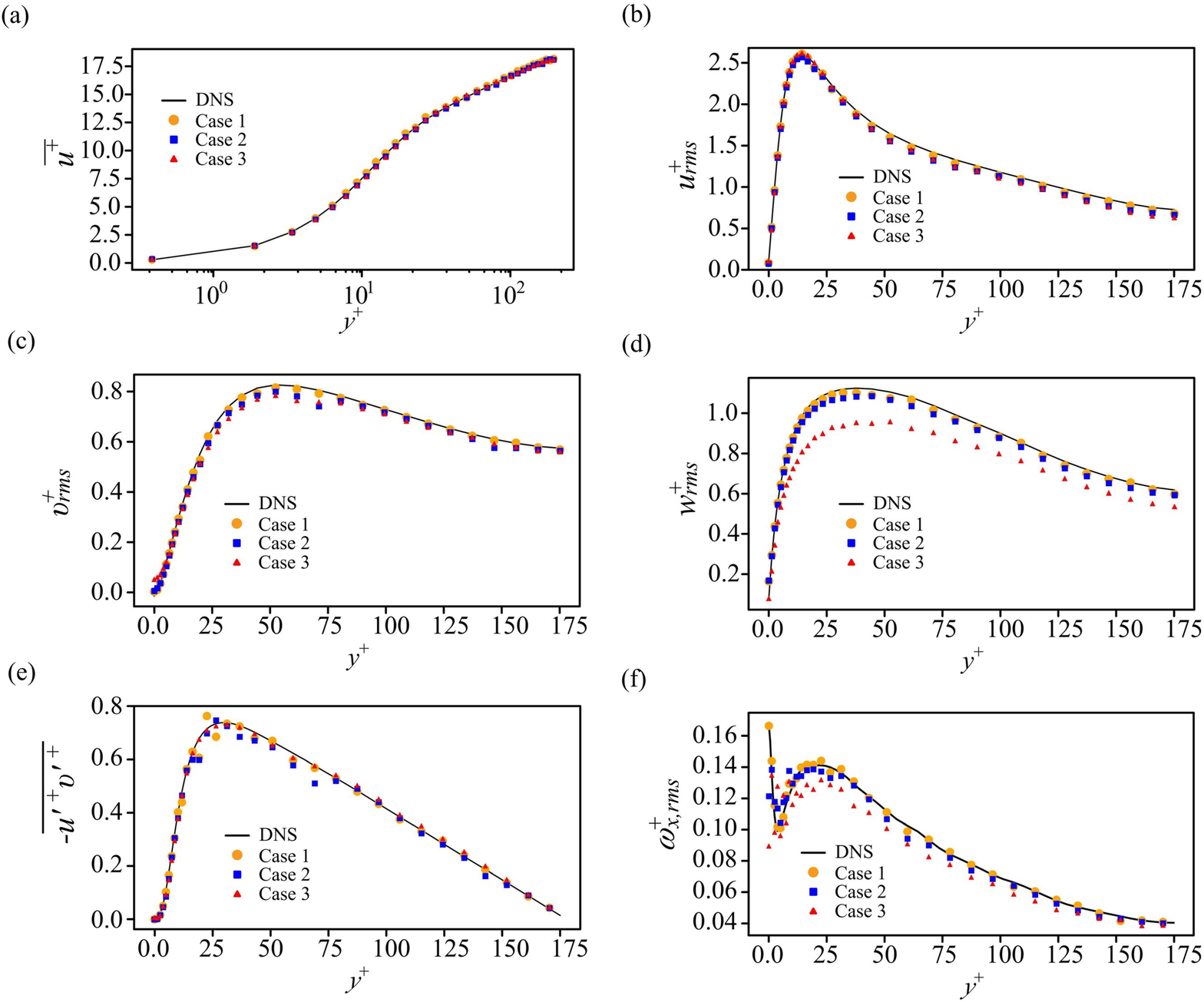}
\caption[]{Turbulence statistics of the reconstructed data for the case of turbulent channel flow: (a) mean streamwise velocity profile, (b) RMS profile of the streamwise velocity, (c) RMS profile of the wall-normal velocity, (d) RMS profile of the spanwise velocity,  (e) mean shear stress profile, and (f) RMS profile of the streamwise vorticity} 
\label{fig:10-Chan-TBL}
\end{figure}

Furthermore, the statistics of 20,000 generated velocity fields corresponding to $t^+ = 22,680$ are compared with the statistics obtained from the DNS results, as shown in Fig.~\ref{fig:10-Chan-TBL}. As can be seen in Fig.~\ref{fig:10-Chan-TBL}(a), the mean streamwise velocity profile for all the three cases of coarseness levels shows excellent agreement with the DNS data obtained within the wall distance range, i.e. the linear viscous sublayer, buffer layer, and logarithmic region. Similarly, the root-mean-square (RMS) profiles of the streamwise and wall-normal velocity components ($u_{rms}^+$ and $\upsilon_{rms}^+$) are also in good agreement with the DNS results for all the coarseness levels, as shown in Fig.~\ref{fig:10-Chan-TBL}(b) and (c). Although the RMS profile of the spanwise velocity component ($w_{rms}^+$) for cases 1 and 2 is in good agreement with the profile obtained using the DNS results, it shows an offset in case 3, as shown in Fig.~\ref{fig:10-Chan-TBL}(d). As mentioned earlier, this can be regarded as the limited information regarding spanwise velocity in case 3. Fig.~\ref{fig:10-Chan-TBL}(e) shows the mean shear stress profile $-\overline{u'^+\upsilon'^+}$. The values are generally in good agreement with the DNS results for all the coarseness levels. Nevertheless, a noticeable scattering  of the values can be seen near $y^+$ $\approx$ 20 - 25 for cases 1 and 2. This can be attributed to the maximum shear stress values that appear in this region which are harder to capture by the model compared with the values that appear in the other regions along the wall distance. Interestingly, the values of  $-\overline{u'^+\upsilon'^+}$  for case 3 show more smooth profile compared with the values for cases 1 and 2 which is contrary to the general expectations considering the previous results. This might be atributed to the under and over-prediction of the streamwise and wall-normal velocity components that can be compensated during the multiplication and averaging processes.

The RMS profiles of the streamwise vorticity ($\omega_{rms}^+$) are shown in Fig.~\ref{fig:10-Chan-TBL}(f). The results for cases 1 and 2 show commendable match with the DNS results. However, case 3 shows less values for most of regions along the wall distance, indicating that the effect of the coarseness level in case 3 is more noticeable compared with those of cases 1 and 2. This would be attributed to a lack of information resulting in impaired reproduction of the streamwise vorticity. 

\begin{figure}
\centering 
\includegraphics[angle=0, trim=0 0 0 0, width=0.8\textwidth]{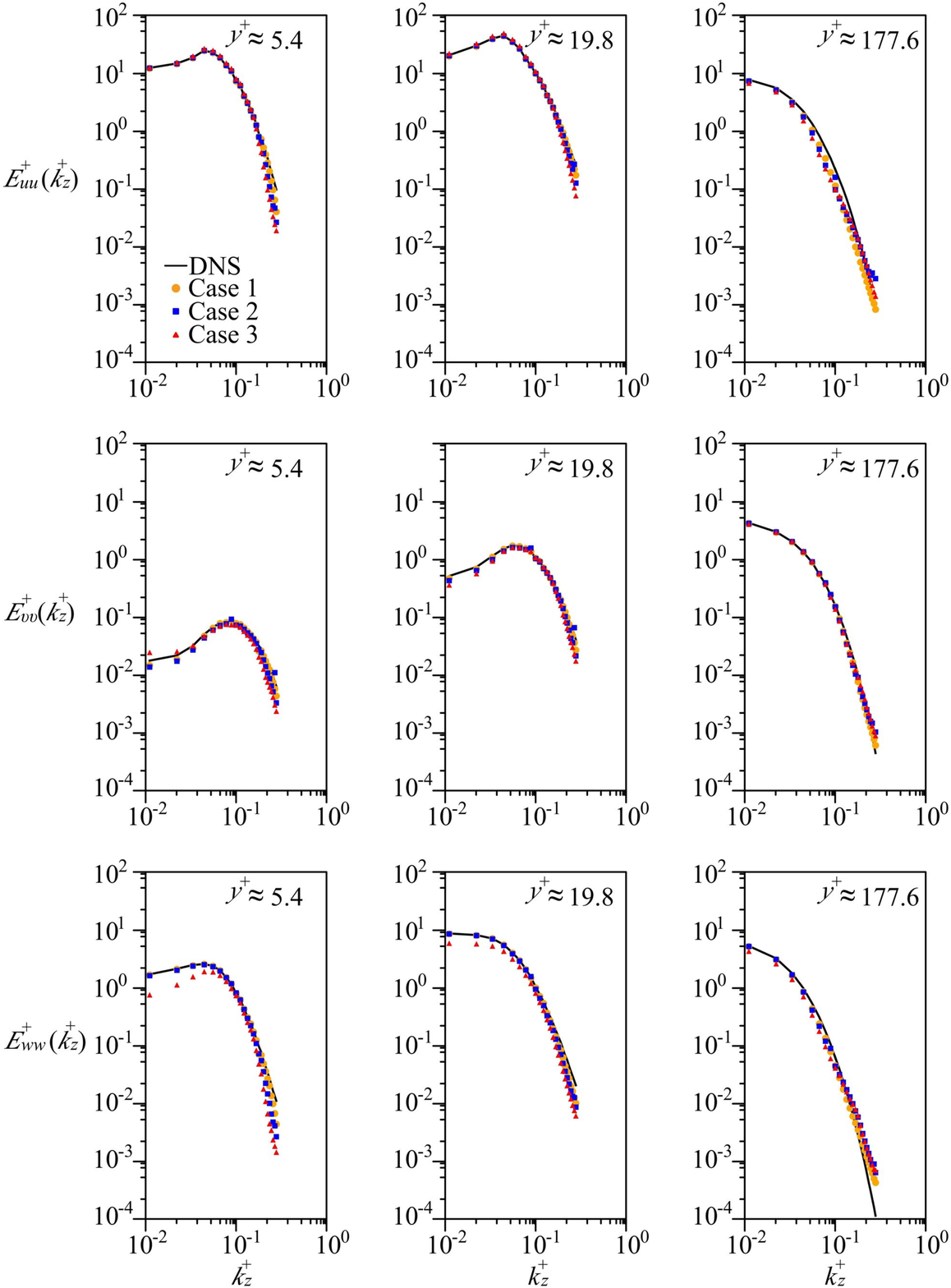}
\caption[]{One-dimensional spanwise energy spectra of the reconstructed velocity components for the case of turbulent channel flow at different wall distances.} 
\label{fig:11-Chan-Spetrum}
\end{figure}

To further investigate the capability of MS-ESRGAN to reconstruct high-resolution velocity fields with realistic behaviour, the one-dimensional spanwise energy spectra for the three cases at different wall distances are shown in Fig.~\ref{fig:11-Chan-Spetrum}. It can be observed from the figure that in all the three cases, the spectral content is reproduced appropriately with a slight deviation at high wavenumbers. These results suggest that the model could successfully reproduce spectra similar to that obtained from the DNS. \par

\begin{figure}
\centering 
\includegraphics[angle=0, trim=0 0 0 0, width=0.8\textwidth]{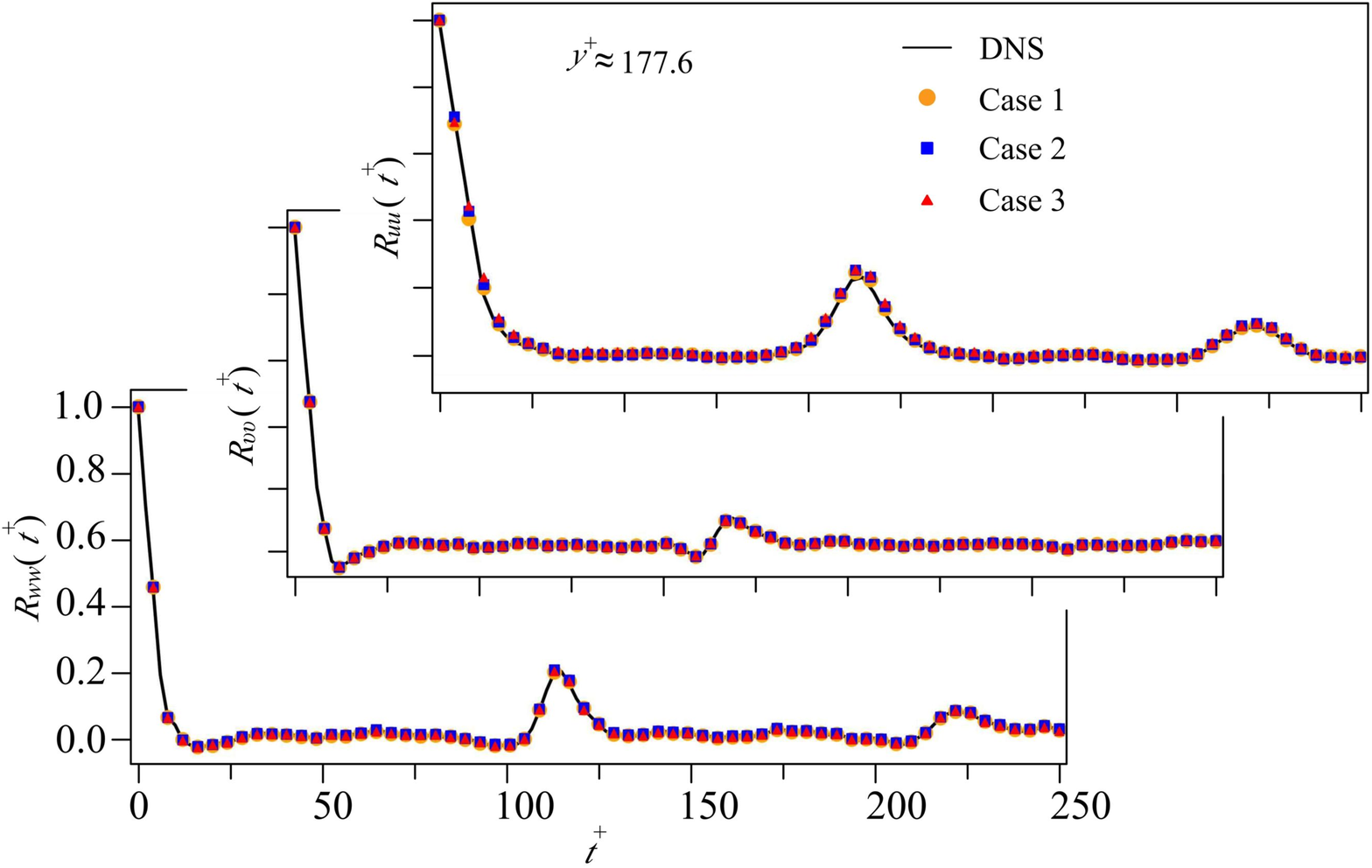}
\caption[]{Time correlations of the reconstructed velocity components for the case of turbulent channel flow.} 
\label{fig:12-Time-Corrln}
\end{figure}

The temporal evolution of the reconstructed snapshots is examined using the time correlation $(R_{ii}(t))$ of each velocity component at $y^+$ $\approx$ 177.6, as shown in Fig.~\ref{fig:12-Time-Corrln}. The results obtained from the reconstructed data are in excellent agreement with the DNS results. Here, the model shows a remarkable ability to reconstruct the velocity data with the same dynamics as the ground truth data.

In summary, the results obtained for the turbulent channel flow case indicate that the proposed MS-ESRGAN, which can successfully reconstruct the velocity fields with high spatial resolution, can reproduce turbulence statistics that are similar to the turbulence statistics of the ground truth data. \par

Our primitive studies revealed that using only $L_G^{Ra}$, $L_{pixel}$, and $L_{perceptual}$ in the generator loss function can result in distorted fluctuations of the velocity components. Furthermore, the reconstructed images showed less sharpness in terms of the flow field details compared with the images reconstructed using the combined loss function. This suggests that using physics-based constraints can remarkably improve the accuracy of the model output.

\subsection{Computational cost}

As a final remark, the computational cost of the proposed MS-ESRGAN is presented in Table~\ref{tab:Table1}. The total number of trainable parameters for both examples in this study is approximately 51 million (45.8 million for $G$ and 5.2 million for $D$). The training of the model on a single GPU machine (Nvidia TITAN RTX) requires approximately 18.8 h for the case of two-dimensional laminar flow around a square cylinder, whereas the training of the turbulent channel flow case requires approximately 65.6 h. This computational cost is required only once to learn how to map the low-resolution flow fields to the high-resolution ones. The reconstruction process of the high-resolution flow fields using the proposed MS-ESRGAN is considered to be computationally inexpensive, as shown at the bottom of Table~\ref{tab:Table1}.

\begin{table}
  \begin{center}
\begin{tabular}{ccc} \hline
&Flow around a square cylinder  ~~~   &   ~~~  Turbulent channel flow \\ \hline
&$G$~~~~~~~~~~~~$D$&$G$~~~~~~~~~~~~$D$ \\ \hline
\makecell{No. of trainable\\ parameters (million)}& 45.8~~~~~~~~~~~~5.2&45.8~~~~~~~~~~~~5.2\\ \hline
Training time(h)&18.8&65.6\\ \hline
Reconstruction time (s)&$1.22\times10^{-2}$&$7.81\times10^{-3}$\\ \hline
\end{tabular}
  \caption{Number of trainable parameters and computational cost of MS-ESRGAN.}
  \label{tab:Table1}
  \end{center}
\end{table}

\section{Conclusions}
In this study, a deep learning-based framework was proposed for the reconstruction of high-resolution turbulent flow fields from spatially limited flow data. We developed an improved version of ESRGAN, i.e. MS-ESRGAN, and applied it to reconstruct the flow fields using distributed points at different levels of coarseness. A combined loss function that includes physics-based loss terms was utilised in $G$ to obtain more realistic results. First, two-dimensional laminar flow around a square cylinder at $Re_d = 100$ simulated using DNS was considered as an illustration of the careful observation of high-resolution flow fields reconstruction using MS-ESRGAN. The model showed a remarkable ability to reconstruct the laminar flow with precise spatial and temporal details even when only minimal spatial flow information was available. The ability of the model to reconstruct wall-bounded turbulence was examined using data from DNS of turbulent channel flow at $Re_\tau = 180$. The model reproduced the instantaneous velocity fields successfully with commendable accuracy for all the three coarseness levels used in the study. Moreover, the turbulence statistics were reproduced appropriately with a slight deviation, which was noticed when very limited spatial information of the velocity fields was provided. The spectra, spatial correlations, and time correlations were also in agreement with the data obtained from the DNS, indicating that the developed model could accurately reconstruct the velocity fields with similar spatial and temporal accuracy as that of the DNS data. In this study, the developed MS-ESRGAN could effectively map flow fields with minimal spatial distribution to a high-resolution ones by utilising the principle of GAN combined with the physics-based loss function. This motivates us to explore more physics-guided deep learning models that can serve as an efficient and inexpensive data-driven methods for recovering high-resolution turbulent flow fields from limited spatial information.

\begin{acknowledgments}
This work was supported by 'Human Resources Program in Energy Technology' of the Korea Institute of Energy Technology Evaluation and Planning (KETEP), granted financial resource from the Ministry of Trade, Industry \& Energy, Republic of Korea (no. 20214000000140). In addition, this work was supported by the National Research Foundation of Korea (NRF) grant funded by the Korea government (MSIP) (no. 2019R1I1A3A01058576). 
\end{acknowledgments}

\section*{Data Availability}
The data that supports the findings of this study are available within this article.

\appendix

\nocite{*}
\bibliography{my-bib}

\providecommand{\noopsort}[1]{}\providecommand{\singleletter}[1]{#1}%
\begin{thebibliography}{31}%
\makeatletter
\providecommand \@ifxundefined [1]{%
 \@ifx{#1\undefined}
}%
\providecommand \@ifnum [1]{%
 \ifnum #1\expandafter \@firstoftwo
 \else \expandafter \@secondoftwo
 \fi
}%
\providecommand \@ifx [1]{%
 \ifx #1\expandafter \@firstoftwo
 \else \expandafter \@secondoftwo
 \fi
}%
\providecommand \natexlab [1]{#1}%
\providecommand \enquote  [1]{``#1''}%
\providecommand \bibnamefont  [1]{#1}%
\providecommand \bibfnamefont [1]{#1}%
\providecommand \citenamefont [1]{#1}%
\providecommand \href@noop [0]{\@secondoftwo}%
\providecommand \href [0]{\begingroup \@sanitize@url \@href}%
\providecommand \@href[1]{\@@startlink{#1}\@@href}%
\providecommand \@@href[1]{\endgroup#1\@@endlink}%
\providecommand \@sanitize@url [0]{\catcode `\\12\catcode `\$12\catcode
  `\&12\catcode `\#12\catcode `\^12\catcode `\_12\catcode `\%12\relax}%
\providecommand \@@startlink[1]{}%
\providecommand \@@endlink[0]{}%
\providecommand \url  [0]{\begingroup\@sanitize@url \@url }%
\providecommand \@url [1]{\endgroup\@href {#1}{\urlprefix }}%
\providecommand \urlprefix  [0]{URL }%
\providecommand \Eprint [0]{\href }%
\providecommand \doibase [0]{https://doi.org/}%
\providecommand \selectlanguage [0]{\@gobble}%
\providecommand \bibinfo  [0]{\@secondoftwo}%
\providecommand \bibfield  [0]{\@secondoftwo}%
\providecommand \translation [1]{[#1]}%
\providecommand \BibitemOpen [0]{}%
\providecommand \bibitemStop [0]{}%
\providecommand \bibitemNoStop [0]{.\EOS\space}%
\providecommand \EOS [0]{\spacefactor3000\relax}%
\providecommand \BibitemShut  [1]{\csname bibitem#1\endcsname}%
\let\auto@bib@innerbib\@empty
\bibitem [{\citenamefont {LeCun}, \citenamefont {Bengio},\ and\ \citenamefont
  {Hinton}(2015)}]{LeCunetal2015}%
  \BibitemOpen
  \bibfield  {author} {\bibinfo {author} {\bibfnamefont {Y.}~\bibnamefont
  {LeCun}}, \bibinfo {author} {\bibfnamefont {Y.}~\bibnamefont {Bengio}},\ and\
  \bibinfo {author} {\bibfnamefont {G.}~\bibnamefont {Hinton}},\ }\bibfield
  {title} {\enquote {\bibinfo {title} {Deep learning},}\ }\href@noop {}
  {\bibfield  {journal} {\bibinfo  {journal} {Nature}\ }\textbf {\bibinfo
  {volume} {521(7553)}},\ \bibinfo {pages} {436--444} (\bibinfo {year}
  {2015})}\BibitemShut {NoStop}%
\bibitem [{\citenamefont {Brunton}, \citenamefont {Noack},\ and\ \citenamefont
  {Koumoutsakos}(2020)}]{Bruntonetal2020}%
  \BibitemOpen
  \bibfield  {author} {\bibinfo {author} {\bibfnamefont {S.~L.}\ \bibnamefont
  {Brunton}}, \bibinfo {author} {\bibfnamefont {B.~R.}\ \bibnamefont {Noack}},\
  and\ \bibinfo {author} {\bibfnamefont {P.}~\bibnamefont {Koumoutsakos}},\
  }\bibfield  {title} {\enquote {\bibinfo {title} {Machine learning for fluid
  mechanics},}\ }\href@noop {} {\bibfield  {journal} {\bibinfo  {journal}
  {Annual Review of Fluid Mechanics}\ }\textbf {\bibinfo {volume} {52(1)}},\
  \bibinfo {pages} {477--508} (\bibinfo {year} {2020})}\BibitemShut {NoStop}%
\bibitem [{\citenamefont {Kutz}(2017)}]{Kutz2017}%
  \BibitemOpen
  \bibfield  {author} {\bibinfo {author} {\bibfnamefont {J.~N.}\ \bibnamefont
  {Kutz}},\ }\bibfield  {title} {\enquote {\bibinfo {title} {Deep learning in
  fluid dynamics},}\ }\href@noop {} {\bibfield  {journal} {\bibinfo  {journal}
  {Journal of Fluid Mechanics}\ }\textbf {\bibinfo {volume} {814}},\ \bibinfo
  {pages} {1--4} (\bibinfo {year} {2017})}\BibitemShut {NoStop}%
\bibitem [{\citenamefont {Duraisamy}\ and\ \citenamefont
  {Xiao}(2019)}]{Duraisamyetal2019}%
  \BibitemOpen
  \bibfield  {author} {\bibinfo {author} {\bibfnamefont {K.}~\bibnamefont
  {Duraisamy}}\ and\ \bibinfo {author} {\bibfnamefont {G.~I.~H.}\ \bibnamefont
  {Xiao}},\ }\bibfield  {title} {\enquote {\bibinfo {title} {Turbulence
  modeling in the age of data},}\ }\href@noop {} {\bibfield  {journal}
  {\bibinfo  {journal} {Annual Review of Fluid Mechanics}\ }\textbf {\bibinfo
  {volume} {51(1)}},\ \bibinfo {pages} {357--377} (\bibinfo {year}
  {2019})}\BibitemShut {NoStop}%
\bibitem [{\citenamefont {Gamahara}\ and\ \citenamefont
  {Hattori}(2017)}]{Gamahara&Hattori2017}%
  \BibitemOpen
  \bibfield  {author} {\bibinfo {author} {\bibfnamefont {M.}~\bibnamefont
  {Gamahara}}\ and\ \bibinfo {author} {\bibfnamefont {Y.}~\bibnamefont
  {Hattori}},\ }\bibfield  {title} {\enquote {\bibinfo {title} {Searching for
  turbulence models by artificial neural network},}\ }\href@noop {} {\bibfield
  {journal} {\bibinfo  {journal} {Physical Review Fluids}\ }\textbf {\bibinfo
  {volume} {2(5)}},\ \bibinfo {pages} {054604} (\bibinfo {year}
  {2017})}\BibitemShut {NoStop}%
\bibitem [{\citenamefont {Ling}, \citenamefont {Kurzawski},\ and\ \citenamefont
  {Templeton}(2016)}]{Lingetal2016}%
  \BibitemOpen
  \bibfield  {author} {\bibinfo {author} {\bibfnamefont {J.}~\bibnamefont
  {Ling}}, \bibinfo {author} {\bibfnamefont {A.}~\bibnamefont {Kurzawski}},\
  and\ \bibinfo {author} {\bibfnamefont {J.}~\bibnamefont {Templeton}},\
  }\bibfield  {title} {\enquote {\bibinfo {title} {Reynolds averaged turbulence
  modelling using deep neural networks with embedded invariance},}\ }\href@noop
  {} {\bibfield  {journal} {\bibinfo  {journal} {Journal of Fluid Mechanics}\
  }\textbf {\bibinfo {volume} {807}},\ \bibinfo {pages} {155--166} (\bibinfo
  {year} {2016})}\BibitemShut {NoStop}%
\bibitem [{\citenamefont {Wang}, \citenamefont {Wu},\ and\ \citenamefont
  {Xiao}(2017)}]{Wangetal2017}%
  \BibitemOpen
  \bibfield  {author} {\bibinfo {author} {\bibfnamefont {J.~X.}\ \bibnamefont
  {Wang}}, \bibinfo {author} {\bibfnamefont {J.~L.}\ \bibnamefont {Wu}},\ and\
  \bibinfo {author} {\bibfnamefont {H.}~\bibnamefont {Xiao}},\ }\bibfield
  {title} {\enquote {\bibinfo {title} {A physics informed machine learning
  approach for reconstructing reynolds stress modeling discrepancies based on
  dns data},}\ }\href@noop {} {\bibfield  {journal} {\bibinfo  {journal}
  {Physical Review Fluids}\ }\textbf {\bibinfo {volume} {2(3)}},\ \bibinfo
  {pages} {034603} (\bibinfo {year} {2017})}\BibitemShut {NoStop}%
\bibitem [{\citenamefont {Lee}\ and\ \citenamefont {You}(2019)}]{Lee&You2019}%
  \BibitemOpen
  \bibfield  {author} {\bibinfo {author} {\bibfnamefont {S.}~\bibnamefont
  {Lee}}\ and\ \bibinfo {author} {\bibfnamefont {D.}~\bibnamefont {You}},\
  }\bibfield  {title} {\enquote {\bibinfo {title} {Data-driven prediction of
  unsteady flow over a circular cylinder using deep learning},}\ }\href@noop {}
  {\bibfield  {journal} {\bibinfo  {journal} {Journal of Fluid Mechanics}\
  }\textbf {\bibinfo {volume} {879}},\ \bibinfo {pages} {217--254} (\bibinfo
  {year} {2019})}\BibitemShut {NoStop}%
\bibitem [{\citenamefont {Srinivasan}\ \emph {et~al.}(2019)\citenamefont
  {Srinivasan}, \citenamefont {Guastoni}, \citenamefont {Azizpour},
  \citenamefont {Schlatter},\ and\ \citenamefont
  {Vinuesa}}]{Srinivasanetal2019}%
  \BibitemOpen
  \bibfield  {author} {\bibinfo {author} {\bibfnamefont {P.~A.}\ \bibnamefont
  {Srinivasan}}, \bibinfo {author} {\bibfnamefont {L.}~\bibnamefont
  {Guastoni}}, \bibinfo {author} {\bibfnamefont {H.}~\bibnamefont {Azizpour}},
  \bibinfo {author} {\bibfnamefont {P.}~\bibnamefont {Schlatter}},\ and\
  \bibinfo {author} {\bibfnamefont {R.}~\bibnamefont {Vinuesa}},\ }\bibfield
  {title} {\enquote {\bibinfo {title} {Predictions of turbulent shear flows
  using deep neural networks},}\ }\href@noop {} {\bibfield  {journal} {\bibinfo
   {journal} {Physical Review Fluids}\ }\textbf {\bibinfo {volume} {4(5)}},\
  \bibinfo {pages} {054603} (\bibinfo {year} {2019})}\BibitemShut {NoStop}%
\bibitem [{\citenamefont {Fan}\ \emph {et~al.}(2020)\citenamefont {Fan},
  \citenamefont {Yang}, \citenamefont {Wang}, \citenamefont {Triantafyllou},\
  and\ \citenamefont {Karniadakis}}]{Fanetal2020}%
  \BibitemOpen
  \bibfield  {author} {\bibinfo {author} {\bibfnamefont {D.}~\bibnamefont
  {Fan}}, \bibinfo {author} {\bibfnamefont {L.}~\bibnamefont {Yang}}, \bibinfo
  {author} {\bibfnamefont {Z.}~\bibnamefont {Wang}}, \bibinfo {author}
  {\bibfnamefont {M.~S.}\ \bibnamefont {Triantafyllou}},\ and\ \bibinfo
  {author} {\bibfnamefont {G.~E.}\ \bibnamefont {Karniadakis}},\ }\bibfield
  {title} {\enquote {\bibinfo {title} {Reinforcement learning for bluff body
  active flow control in experiments and simulations},}\ }\href@noop {}
  {\bibfield  {journal} {\bibinfo  {journal} {Proceedings of the National
  Academy of Sciences}\ }\textbf {\bibinfo {volume} {117(42)}},\ \bibinfo
  {pages} {26091--26098.} (\bibinfo {year} {2020})}\BibitemShut {NoStop}%
\bibitem [{\citenamefont {Rabault}\ \emph {et~al.}(2019)\citenamefont
  {Rabault}, \citenamefont {Kuchta}, \citenamefont {Jensen}, \citenamefont
  {R{\'e}glade},\ and\ \citenamefont {Cerardi}}]{Rabaultetal2019}%
  \BibitemOpen
  \bibfield  {author} {\bibinfo {author} {\bibfnamefont {J.}~\bibnamefont
  {Rabault}}, \bibinfo {author} {\bibfnamefont {M.}~\bibnamefont {Kuchta}},
  \bibinfo {author} {\bibfnamefont {A.}~\bibnamefont {Jensen}}, \bibinfo
  {author} {\bibfnamefont {U.}~\bibnamefont {R{\'e}glade}},\ and\ \bibinfo
  {author} {\bibfnamefont {N.}~\bibnamefont {Cerardi}},\ }\bibfield  {title}
  {\enquote {\bibinfo {title} {Artificial neural networks trained through deep
  reinforcement learning discover control strategies for active flow
  control},}\ }\href@noop {} {\bibfield  {journal} {\bibinfo  {journal}
  {Journal of Fluid Mechanics}\ }\textbf {\bibinfo {volume} {865}},\ \bibinfo
  {pages} {281--302} (\bibinfo {year} {2019})}\BibitemShut {NoStop}%
\bibitem [{\citenamefont {Bashir}\ \emph {et~al.}(2021)\citenamefont {Bashir},
  \citenamefont {Wang}, \citenamefont {Khan},\ and\ \citenamefont
  {Niu.}}]{Bashiretal2021}%
  \BibitemOpen
  \bibfield  {author} {\bibinfo {author} {\bibfnamefont {S.~M.~A.}\
  \bibnamefont {Bashir}}, \bibinfo {author} {\bibfnamefont {Y.}~\bibnamefont
  {Wang}}, \bibinfo {author} {\bibfnamefont {M.}~\bibnamefont {Khan}},\ and\
  \bibinfo {author} {\bibfnamefont {Y.}~\bibnamefont {Niu.}},\ }\bibfield
  {title} {\enquote {\bibinfo {title} {A comprehensive review of deep
  learning-based single image super-resolution},}\ }\href@noop {} {\bibfield
  {journal} {\bibinfo  {journal} {PeerJ Computer Science}\ }\textbf {\bibinfo
  {volume} {7}},\ \bibinfo {pages} {e621} (\bibinfo {year} {2021})}\BibitemShut
  {NoStop}%
\bibitem [{\citenamefont {Keys}(1981)}]{Keys1981}%
  \BibitemOpen
  \bibfield  {author} {\bibinfo {author} {\bibfnamefont {R.}~\bibnamefont
  {Keys}},\ }\bibfield  {title} {\enquote {\bibinfo {title} {Cubic convolution
  interpolation for digital image processing},}\ }\href@noop {} {\bibfield
  {journal} {\bibinfo  {journal} {IEEE Transactions on Acoustics, Speech, and
  Signal Processing}\ }\textbf {\bibinfo {volume} {29(6)}},\ \bibinfo {pages}
  {1153--1160} (\bibinfo {year} {1981})}\BibitemShut {NoStop}%
\bibitem [{\citenamefont {Fukami}, \citenamefont {Fukagata},\ and\
  \citenamefont {Taira}(2019)}]{Fukami2019}%
  \BibitemOpen
  \bibfield  {author} {\bibinfo {author} {\bibfnamefont {K.}~\bibnamefont
  {Fukami}}, \bibinfo {author} {\bibfnamefont {K.}~\bibnamefont {Fukagata}},\
  and\ \bibinfo {author} {\bibfnamefont {K.}~\bibnamefont {Taira}},\ }\bibfield
   {title} {\enquote {\bibinfo {title} {Super-resolution reconstruction of
  turbulent flows with machine learning},}\ }\href@noop {} {\bibfield
  {journal} {\bibinfo  {journal} {Journal of Fluid Mechanics}\ }\textbf
  {\bibinfo {volume} {870}},\ \bibinfo {pages} {106--120} (\bibinfo {year}
  {2019})}\BibitemShut {NoStop}%
\bibitem [{\citenamefont {Fukami}, \citenamefont {Fukagata},\ and\
  \citenamefont {Taira}(2021)}]{Fukami2021}%
  \BibitemOpen
  \bibfield  {author} {\bibinfo {author} {\bibfnamefont {K.}~\bibnamefont
  {Fukami}}, \bibinfo {author} {\bibfnamefont {K.}~\bibnamefont {Fukagata}},\
  and\ \bibinfo {author} {\bibfnamefont {K.}~\bibnamefont {Taira}},\ }\bibfield
   {title} {\enquote {\bibinfo {title} {Machine-learning-based spatio-temporal
  super resolution reconstruction of turbulent flows},}\ }\href@noop {}
  {\bibfield  {journal} {\bibinfo  {journal} {Journal of Fluid Mechanics}\
  }\textbf {\bibinfo {volume} {909}} (\bibinfo {year} {2021})}\BibitemShut
  {NoStop}%
\bibitem [{\citenamefont {Onishi}, \citenamefont {Sugiyama},\ and\
  \citenamefont {Matsuda}(2019)}]{Onishietal2019}%
  \BibitemOpen
  \bibfield  {author} {\bibinfo {author} {\bibfnamefont {R.}~\bibnamefont
  {Onishi}}, \bibinfo {author} {\bibfnamefont {D.}~\bibnamefont {Sugiyama}},\
  and\ \bibinfo {author} {\bibfnamefont {K.}~\bibnamefont {Matsuda}},\
  }\bibfield  {title} {\enquote {\bibinfo {title} {Super-resolution simulation
  for real-time prediction of urban micrometerology},}\ }\href@noop {}
  {\bibfield  {journal} {\bibinfo  {journal} {Scientific Online Letters on the
  Atmosphere}\ }\textbf {\bibinfo {volume} {15}},\ \bibinfo {pages} {178--182}
  (\bibinfo {year} {2019})}\BibitemShut {NoStop}%
\bibitem [{\citenamefont {Liu}\ \emph {et~al.}(2020)\citenamefont {Liu},
  \citenamefont {Tang}, \citenamefont {Huang},\ and\ \citenamefont
  {Lu}}]{Liuetal2020}%
  \BibitemOpen
  \bibfield  {author} {\bibinfo {author} {\bibfnamefont {B.}~\bibnamefont
  {Liu}}, \bibinfo {author} {\bibfnamefont {J.}~\bibnamefont {Tang}}, \bibinfo
  {author} {\bibfnamefont {H.}~\bibnamefont {Huang}},\ and\ \bibinfo {author}
  {\bibfnamefont {X.~Y.}\ \bibnamefont {Lu}},\ }\bibfield  {title} {\enquote
  {\bibinfo {title} {Deep learning methods for super-resolution reconstruction
  of turbulent flows},}\ }\href@noop {} {\bibfield  {journal} {\bibinfo
  {journal} {Physics of Fluids}\ }\textbf {\bibinfo {volume} {32(2)}},\
  \bibinfo {pages} {025105} (\bibinfo {year} {2020})}\BibitemShut {NoStop}%
\bibitem [{\citenamefont {Kim}\ \emph {et~al.}(2021)\citenamefont {Kim},
  \citenamefont {Kim}, \citenamefont {Won},\ and\ \citenamefont
  {Lee}}]{Kimetal2021}%
  \BibitemOpen
  \bibfield  {author} {\bibinfo {author} {\bibfnamefont {H.}~\bibnamefont
  {Kim}}, \bibinfo {author} {\bibfnamefont {J.}~\bibnamefont {Kim}}, \bibinfo
  {author} {\bibfnamefont {S.}~\bibnamefont {Won}},\ and\ \bibinfo {author}
  {\bibfnamefont {C.}~\bibnamefont {Lee}},\ }\bibfield  {title} {\enquote
  {\bibinfo {title} {Unsupervised deep learning for super-resolution
  reconstruction of turbulence},}\ }\href@noop {} {\bibfield  {journal}
  {\bibinfo  {journal} {Journal of Fluid Mechanics}\ }\textbf {\bibinfo
  {volume} {910}} (\bibinfo {year} {2021})}\BibitemShut {NoStop}%
\bibitem [{\citenamefont {Zhu}\ \emph {et~al.}(2017)\citenamefont {Zhu},
  \citenamefont {Park}, \citenamefont {Isola},\ and\ \citenamefont
  {Efros}}]{Zhuetal2017}%
  \BibitemOpen
  \bibfield  {author} {\bibinfo {author} {\bibfnamefont {J.~Y.}\ \bibnamefont
  {Zhu}}, \bibinfo {author} {\bibfnamefont {T.}~\bibnamefont {Park}}, \bibinfo
  {author} {\bibfnamefont {P.}~\bibnamefont {Isola}},\ and\ \bibinfo {author}
  {\bibfnamefont {A.~A.}\ \bibnamefont {Efros}},\ }\bibfield  {title} {\enquote
  {\bibinfo {title} {Unpaired image-to-image translation using cycle-consistent
  adversarial networks},}\ }\href@noop {} {\bibfield  {journal} {\bibinfo
  {journal} {2017 IEEE International Conference on Computer Vision (ICCV)}\ ,\
  \bibinfo {pages} {2242–2251}} (\bibinfo {year} {2017})}\BibitemShut
  {NoStop}%
\bibitem [{\citenamefont {Deng}\ \emph {et~al.}(2019)\citenamefont {Deng},
  \citenamefont {He}, \citenamefont {Liu},\ and\ \citenamefont
  {Kim}}]{Dengetal2019}%
  \BibitemOpen
  \bibfield  {author} {\bibinfo {author} {\bibfnamefont {Z.}~\bibnamefont
  {Deng}}, \bibinfo {author} {\bibfnamefont {C.}~\bibnamefont {He}}, \bibinfo
  {author} {\bibfnamefont {Y.}~\bibnamefont {Liu}},\ and\ \bibinfo {author}
  {\bibfnamefont {K.~C.}\ \bibnamefont {Kim}},\ }\bibfield  {title} {\enquote
  {\bibinfo {title} {Super-resolution reconstruction of turbulent velocity
  fields using a generative adversarial network-based artificial intelligence
  framework},}\ }\href@noop {} {\bibfield  {journal} {\bibinfo  {journal}
  {Physics of Fluids}\ }\textbf {\bibinfo {volume} {31(12)}},\ \bibinfo {pages}
  {125111} (\bibinfo {year} {2019})}\BibitemShut {NoStop}%
\bibitem [{\citenamefont {Ledig}\ \emph {et~al.}(2017)\citenamefont {Ledig},
  \citenamefont {Theis}, \citenamefont {Huszar}, \citenamefont {Caballero},
  \citenamefont {Cunningham}, \citenamefont {Acosta}, \citenamefont {Aitken},
  \citenamefont {Tejan}, \citenamefont {Totz}, \citenamefont {Wang},\ and\
  \citenamefont {Shi}}]{Ledigetal2017}%
  \BibitemOpen
  \bibfield  {author} {\bibinfo {author} {\bibfnamefont {C.}~\bibnamefont
  {Ledig}}, \bibinfo {author} {\bibfnamefont {L.}~\bibnamefont {Theis}},
  \bibinfo {author} {\bibfnamefont {F.}~\bibnamefont {Huszar}}, \bibinfo
  {author} {\bibfnamefont {J.}~\bibnamefont {Caballero}}, \bibinfo {author}
  {\bibfnamefont {A.}~\bibnamefont {Cunningham}}, \bibinfo {author}
  {\bibfnamefont {A.}~\bibnamefont {Acosta}}, \bibinfo {author} {\bibfnamefont
  {A.}~\bibnamefont {Aitken}}, \bibinfo {author} {\bibfnamefont
  {A.}~\bibnamefont {Tejan}}, \bibinfo {author} {\bibfnamefont
  {J.}~\bibnamefont {Totz}}, \bibinfo {author} {\bibfnamefont {Z.}~\bibnamefont
  {Wang}},\ and\ \bibinfo {author} {\bibfnamefont {W.}~\bibnamefont {Shi}},\
  }\bibfield  {title} {\enquote {\bibinfo {title} {Photo-realistic single image
  super-resolution using a generative adversarial network},}\ }\href@noop {}
  {\bibfield  {journal} {\bibinfo  {journal} {ArXiv:1609.04802 [Cs, Stat]}\ }
  (\bibinfo {year} {2017})}\BibitemShut {NoStop}%
\bibitem [{\citenamefont {Wang}\ \emph {et~al.}(2018)\citenamefont {Wang},
  \citenamefont {Yu}, \citenamefont {Wu}, \citenamefont {Gu}, \citenamefont
  {Liu}, \citenamefont {Dong}, \citenamefont {Loy}, \citenamefont {Qiao},\ and\
  \citenamefont {Tang}}]{Wangetal2018}%
  \BibitemOpen
  \bibfield  {author} {\bibinfo {author} {\bibfnamefont {X.}~\bibnamefont
  {Wang}}, \bibinfo {author} {\bibfnamefont {K.}~\bibnamefont {Yu}}, \bibinfo
  {author} {\bibfnamefont {S.}~\bibnamefont {Wu}}, \bibinfo {author}
  {\bibfnamefont {J.}~\bibnamefont {Gu}}, \bibinfo {author} {\bibfnamefont
  {Y.}~\bibnamefont {Liu}}, \bibinfo {author} {\bibfnamefont {C.}~\bibnamefont
  {Dong}}, \bibinfo {author} {\bibfnamefont {C.~C.}\ \bibnamefont {Loy}},
  \bibinfo {author} {\bibfnamefont {Y.}~\bibnamefont {Qiao}},\ and\ \bibinfo
  {author} {\bibfnamefont {X.}~\bibnamefont {Tang}},\ }\bibfield  {title}
  {\enquote {\bibinfo {title} {Esrgan: Enhanced super-resolution generative
  adversarial networks},}\ }\href@noop {} {\bibfield  {journal} {\bibinfo
  {journal} {rXiv:1809.00219 [Cs]}\ } (\bibinfo {year} {2018})}\BibitemShut
  {NoStop}%
\bibitem [{\citenamefont {Cai}\ \emph {et~al.}(2019)\citenamefont {Cai},
  \citenamefont {Zhou}, \citenamefont {Xu},\ and\ \citenamefont
  {Gao}}]{Caietal2019}%
  \BibitemOpen
  \bibfield  {author} {\bibinfo {author} {\bibfnamefont {S.}~\bibnamefont
  {Cai}}, \bibinfo {author} {\bibfnamefont {S.}~\bibnamefont {Zhou}}, \bibinfo
  {author} {\bibfnamefont {C.}~\bibnamefont {Xu}},\ and\ \bibinfo {author}
  {\bibfnamefont {Q.}~\bibnamefont {Gao}},\ }\bibfield  {title} {\enquote
  {\bibinfo {title} {Dense motion estimation of particale images via a
  convolutional neural network},}\ }\href@noop {} {\bibfield  {journal}
  {\bibinfo  {journal} {Exp Fluids}\ }\textbf {\bibinfo {volume} {60}},\
  \bibinfo {pages} {73} (\bibinfo {year} {2019})}\BibitemShut {NoStop}%
\bibitem [{\citenamefont {Morimoto}, \citenamefont {Fukami},\ and\
  \citenamefont {Fukagata}(2020)}]{Morimotoetal2020}%
  \BibitemOpen
  \bibfield  {author} {\bibinfo {author} {\bibfnamefont {M.}~\bibnamefont
  {Morimoto}}, \bibinfo {author} {\bibfnamefont {K.}~\bibnamefont {Fukami}},\
  and\ \bibinfo {author} {\bibfnamefont {K.}~\bibnamefont {Fukagata}},\
  }\bibfield  {title} {\enquote {\bibinfo {title} {Experimental velocity data
  estimation for imperfect particle images using machine learning},}\
  }\href@noop {} {\bibfield  {journal} {\bibinfo  {journal} {ArXiv:2005.00756
  [Physics]}\ } (\bibinfo {year} {2020})}\BibitemShut {NoStop}%
\bibitem [{\citenamefont {Goodfellow}\ \emph {et~al.}(2014)\citenamefont
  {Goodfellow}, \citenamefont {Pouget-Abadie}, \citenamefont {Mirza},
  \citenamefont {Xu}, \citenamefont {Warde-Farley}, \citenamefont {Ozair},
  \citenamefont {Courville},\ and\ \citenamefont
  {Y.Bengio}}]{Goodfellowetal2014}%
  \BibitemOpen
  \bibfield  {author} {\bibinfo {author} {\bibfnamefont {I.}~\bibnamefont
  {Goodfellow}}, \bibinfo {author} {\bibfnamefont {J.}~\bibnamefont
  {Pouget-Abadie}}, \bibinfo {author} {\bibfnamefont {M.}~\bibnamefont
  {Mirza}}, \bibinfo {author} {\bibfnamefont {B.}~\bibnamefont {Xu}}, \bibinfo
  {author} {\bibfnamefont {D.}~\bibnamefont {Warde-Farley}}, \bibinfo {author}
  {\bibfnamefont {S.}~\bibnamefont {Ozair}}, \bibinfo {author} {\bibfnamefont
  {A.}~\bibnamefont {Courville}},\ and\ \bibinfo {author} {\bibnamefont
  {Y.Bengio}},\ }\bibfield  {title} {\enquote {\bibinfo {title} {Generative
  adversarial nets},}\ }\href@noop {} {\bibfield  {journal} {\bibinfo
  {journal} {Advances in Neural Information Processing Systems}\ }\textbf
  {\bibinfo {volume} {27}} (\bibinfo {year} {2014})}\BibitemShut {NoStop}%
\bibitem [{\citenamefont {Mirza}\ and\ \citenamefont
  {Osindero}(2014)}]{Mirza&Osindero2014}%
  \BibitemOpen
  \bibfield  {author} {\bibinfo {author} {\bibfnamefont {M.}~\bibnamefont
  {Mirza}}\ and\ \bibinfo {author} {\bibfnamefont {S.}~\bibnamefont
  {Osindero}},\ }\bibfield  {title} {\enquote {\bibinfo {title} {Conditional
  generative adversarial nets},}\ }\href@noop {} {\bibfield  {journal}
  {\bibinfo  {journal} {https://arxiv.org/abs/1411.1784v1}\ } (\bibinfo {year}
  {2014})}\BibitemShut {NoStop}%
\bibitem [{\citenamefont {relativistic discriminator: A key element missing
  from~standard GAN}(2018)}]{Jolicoeur-Martineau2018}%
  \BibitemOpen
  \bibfield  {author} {\bibinfo {author} {\bibfnamefont {T.}~\bibnamefont
  {relativistic discriminator: A key element missing from~standard GAN}},\
  }\bibfield  {title} {\enquote {\bibinfo {title} {A synthetic-eddy-method for
  generating inflow conditions for large-eddy simulations.}}\ }\href@noop {}
  {\bibfield  {journal} {\bibinfo  {journal} {ArXiv:1807.00734 [Cs, Stat]}\ }
  (\bibinfo {year} {2018})}\BibitemShut {NoStop}%
\bibitem [{\citenamefont {K.Simonyan}\ and\ \citenamefont
  {Zisserman}(2015)}]{Simonyan&Zisserman2015}%
  \BibitemOpen
  \bibfield  {author} {\bibinfo {author} {\bibnamefont {K.Simonyan}}\ and\
  \bibinfo {author} {\bibfnamefont {A.}~\bibnamefont {Zisserman}},\ }\bibfield
  {title} {\enquote {\bibinfo {title} {Very deep convolutional networks for
  large-scale image recognition},}\ }\href@noop {} {\bibfield  {journal}
  {\bibinfo  {journal} {ArXiv:1409.1556 [Cs]}\ } (\bibinfo {year}
  {2015})}\BibitemShut {NoStop}%
\bibitem [{\citenamefont {Kingma}\ and\ \citenamefont
  {Ba}(2017)}]{Kingma&Ba2017}%
  \BibitemOpen
  \bibfield  {author} {\bibinfo {author} {\bibfnamefont {D.}~\bibnamefont
  {Kingma}}\ and\ \bibinfo {author} {\bibfnamefont {J.}~\bibnamefont {Ba}},\
  }\bibfield  {title} {\enquote {\bibinfo {title} {Adam: A method for
  stochastic optimization},}\ }\href@noop {} {\bibfield  {journal} {\bibinfo
  {journal} {ArXiv:1412.6980 [Cs]}\ } (\bibinfo {year} {2017})}\BibitemShut
  {NoStop}%
\bibitem [{\citenamefont {Anzai}\ \emph {et~al.}(2017)\citenamefont {Anzai},
  \citenamefont {Fukagata}, \citenamefont {Meliga}, \citenamefont {Boujo},\
  and\ \citenamefont {Gallaire}}]{Anzaietal2017}%
  \BibitemOpen
  \bibfield  {author} {\bibinfo {author} {\bibfnamefont {Y.}~\bibnamefont
  {Anzai}}, \bibinfo {author} {\bibfnamefont {K.}~\bibnamefont {Fukagata}},
  \bibinfo {author} {\bibfnamefont {P.}~\bibnamefont {Meliga}}, \bibinfo
  {author} {\bibfnamefont {E.}~\bibnamefont {Boujo}},\ and\ \bibinfo {author}
  {\bibfnamefont {F.}~\bibnamefont {Gallaire}},\ }\bibfield  {title} {\enquote
  {\bibinfo {title} {Numerical simulation and sensitivity analysis of a
  low-reynolds-number flow around a square cylinder controlled using plasma
  actuators},}\ }\href@noop {} {\bibfield  {journal} {\bibinfo  {journal}
  {Phys. Rev. Fluids}\ }\textbf {\bibinfo {volume} {2 (4)}},\ \bibinfo {pages}
  {043901} (\bibinfo {year} {2017})}\BibitemShut {NoStop}%
\bibitem [{\citenamefont {Moser}, \citenamefont {Kim},\ and\ \citenamefont
  {Mansour}(1999)}]{Moseretal1999}%
  \BibitemOpen
  \bibfield  {author} {\bibinfo {author} {\bibfnamefont {R.~D.}\ \bibnamefont
  {Moser}}, \bibinfo {author} {\bibfnamefont {J.}~\bibnamefont {Kim}},\ and\
  \bibinfo {author} {\bibfnamefont {N.~N.}\ \bibnamefont {Mansour}},\
  }\bibfield  {title} {\enquote {\bibinfo {title} {Direct numerical simulation
  of turbulent channel flow up to $re_\tau=590$},}\ }\href@noop {} {\bibfield
  {journal} {\bibinfo  {journal} {Physics of Fluids}\ }\textbf {\bibinfo
  {volume} {11(4)}},\ \bibinfo {pages} {943--945} (\bibinfo {year}
  {1999})}\BibitemShut {NoStop}%
\end{thebibliography}%

\end{document}